\documentclass[final,5p,times,twocolumn]{elsarticle}  

\usepackage{graphicx}
\usepackage{epsfig}
\usepackage{epstopdf}
\usepackage{subfig}

\usepackage{amssymb}
\usepackage{amsthm}
\usepackage{amsmath}
\usepackage{setspace}
\usepackage{lineno}
\usepackage{hyperref}
\usepackage{verbatim}
\usepackage{filecontents}
\usepackage[all]{hypcap}

\usepackage{color}

\makeatother
    \makeatletter
    \def\ps@pprintTitle{%
       \let\@oddhead\@empty
       \let\@evenhead\@empty
       \let\@oddfoot\@empty
       \let\@evenfoot\@oddfoot
    }
    \makeatother

\begin{document}



\begin{frontmatter}



\title{Design and Performance of an Interferometric Trigger Array \\
for Radio Detection of High-Energy Neutrinos}
\author[OSUPhysics] {P. Allison} 
\author[CHIBA]{S. Archambault} 
\author[UMD]{R. Bard} 
\author[OSUPhysics,OSUAstronomy]{J. J. Beatty} 
\author[UW]{M. Beheler-Amass} 
\author[KU,NRNU]{D. Z. Besson} 
\author[UW]{M. Beydler} 
\author[UCedg]{M.~Bogdan} 
\author[NTU]{C.-C.~Chen} 
\author[NTU]{C.-H. Chen} 
\author[NTU]{P. Chen} 
\author[OSUPhysics]{B. A. Clark}
\author[UNL]{A. Clough} 
\author[OSUPhysics]{A. Connolly}
\author[UCL]{L. Cremonesi} 
\author[UCL]{J. Davies} 
\author[UC]{C.~Deaconu} 
\author[UW]{M.~A.~DuVernois} 
\author[UMD]{E. Friedman} 
\author[Whittier]{J. Hanson} 
\author[UW]{K. Hanson} 
\author[UW]{J. Haugen} 
\author[UMD]{K. D. Hoffman} 
\author[UW]{B. Hokanson-Fasig} 
\author[OSUPhysics]{E. Hong} 
\author[NTU]{S.-Y.~Hsu} 
\author[NTU]{L.~Hu} 
\author[NTU]{J.-J. Huang} 
\author[NTU]{M.-H. Huang} 
\author[UC]{K. Hughes} 
\author[CHIBA]{A. Ishihara} 
\author[UW]{A. Karle} 
\author[UW]{J. L. Kelley} 
\author[UW]{R. Khandelwal} 
\author[CHIBA]{M. Kim} 
\author[UNL]{I.~Kravchenko} 
\author[UNL]{J. Kruse} 
\author[CHIBA]{K. Kurusu} 
\author[Weizmann]{H. Landsman} 
\author[KU] {U. A. Latif} 
\author[UW]{A. Laundrie} 
\author[NTU]{C.-J. Li} 
\author[NTU]{T. C. Liu} 
\author[UW]{M.-Y. Lu} 
\author[UC]{A.~Ludwig} 
\author[CHIBA]{K.~Mase} 
\author[UW]{T. Meures} 
\author[NTU]{J. Nam} 
\author[UCL]{R. J. Nichol} 
\author[Weizmann]{G. Nir} 
\author[UC]{E. Oberla\corref{cor1}} 
\ead{ejo@uchicago.edu}
\author[UW]{A. \'{O}Murchadha} 
\author[UD]{Y.~Pan} 
\author[Otterbein]{C.~Pfendner}
\author[UC]{M. Ransom} 
\author[KU]{K.~Ratzlaff} 
\author[UD]{J. Roth} 
\author[UW]{P. Sandstrom} 
\author[UD]{D. Seckel} 
\author[NTU]{Y.-S. Shiao} 
\author[UNL]{A. Shultz} 
\author[UC]{D.~Smith} 
\author[UMD]{M.~Song} 
\author[Moscow]{M. Sullivan}
\author[UMD]{J. Touart} 
\author[UC,UCedg]{A.~G.~Vieregg} 
\author[NTU]{M.-Z.~Wang} 
\author[NTU]{S.-H. Wang} 
\author[UC]{K.~Wei} 
\author[CalPoly]{S. A. Wissel} 
\author[CHIBA]{S.~Yoshida} 
\author[KU]{R. Young} 

\address[OSUPhysics] {Dept. of Physics and Center for Cosmology and AstroParticle Physics, The Ohio State University, Columbus, OH, USA}
\address[CHIBA]{Dept. of Physics, Chiba University, Chiba, Japan}
\address[UMD]{Dept. of Physics, University of Maryland, College Park, MD, USA}
\address[OSUAstronomy] {Dept. of Astronomy, The Ohio State University, Columbus, OH USA}
\address[UW]{Dept. of Physics and Wisconsin IceCube Particle Astrophysics Center, University of Wisconsin, Madison, WI, USA}
\address[KU]{Dept. of Physics and Astronomy and Instrumentation Design Laboratory, University of Kansas, Lawrence, KS, USA}
\address[NRNU]{National Research Nuclear University, Moscow Engineering Physics Institute, Moscow, Russia}
\address[UCedg]{Enrico Fermi Institute, The University of Chicago, Chicago, IL, USA}
\address[NTU]{Dept.~of~Physics,~Grad.~Inst.~of Astrophys.,~Leung Center for Cosmology and Particle Astrophys., National Taiwan Univ., Taipei, Tawian}
\address[UNL]{Dept. of Physics and Astronomy, University of Nebraska-Lincoln, Lincoln, NE, USA}

\address[UCL]{Dept. of Physics and Astronomy, University College London, London, UK}
\address[UC]{Dept. of Physics and Kavli Institute for Cosmological Physics, The University of Chicago, Chicago, IL, USA}
\address[Whittier]{Dept. of Physics and Astronomy, Whittier College, Whittier, CA, USA}
\address[Weizmann]{Weizmann Institute of Science, Rehovot, Israel}
\address[UD]{Dept. of  Physics  and  Astronomy,  University of Delaware, Newark, DE, USA}
\address[Otterbein]{Physics Dept., Otterbein University, Westerville, OH, USA}
\address[Moscow]{Moscow Engineering and Physics Institute, Moscow, Russia}
\address[CalPoly]{Dept. of Physics, California Polytechnic State University, San Luis Obispo, CA, USA}

\begin{abstract}

Ultra-high energy neutrinos are detectable through impulsive radio 
signals generated through interactions in dense media, such as ice. 
Subsurface in-ice radio arrays are a promising way to advance the
observation and measurement of astrophysical high-energy neutrinos 
with energies above those discovered by the IceCube detector ($\geq$~1~PeV)
as well as cosmogenic neutrinos created in the GZK process ($\geq$~100~PeV). 
Here we describe the {\it NuPhase} detector, which is 
a compact receiving array of low-gain antennas deployed 185~m deep in glacial 
ice near the South Pole. Signals from the antennas are 
digitized and coherently summed into multiple beams 
to form a low-threshold interferometric phased array trigger for radio impulses.
The NuPhase detector was installed 
at an Askaryan Radio Array (ARA) station during the 2017/18 Austral summer season. 
{\it In situ} measurements with an impulsive, 
point-source calibration instrument show a 50\%
trigger efficiency on impulses with voltage signal-to-noise 
ratios (SNR) of $\le$2.0,  a factor of $\sim$1.8 improvement in SNR over the 
standard ARA combinatoric trigger. 
Hardware-level simulations, validated with {\it in situ} measurements, predict
a trigger threshold of an SNR as low as 1.6 for neutrino interactions that are in the far field of the array.  
With the already-achieved NuPhase trigger performance included in ARASim, a detector simulation for the ARA experiment,
we find the trigger-level effective detector volume is increased by a factor of 1.8 
at neutrino energies between 10 and 100~PeV compared to the 
currently used ARA combinatoric trigger. 
We also discuss an achievable near term path toward lowering the 
trigger threshold further to an SNR of 1.0, which would increase the effective single-station
volume by more than a factor of 3 in the same range of neutrino energies.



\end{abstract}



\cortext[cor1]{Corresponding Author}

\end{frontmatter}



\section{Introduction}
\label{sec:intro}
In recent years high-energy neutrinos ($>$~0.1~PeV) of 
astrophysical origin have been discovered by the
IceCube experiment~\cite{icecube2013, icecubenorthern}. 
Using a dataset containing upgoing muon (track-like) events, IceCube
shows that these data are well described by a relatively hard spectrum power-law ($E^{-2.1}$),
disfavoring flux models with an exponential energy cut-off~\cite{icecubeupgoing}.
A recent multi-messenger observation of a $\sim$0.3~PeV neutrino from the direction
of a gamma-ray flaring blazar provides a clue to progenitors of these neutrinos~\cite{icecubeblazar}.
At higher energies, ultra-high energy 
neutrinos ($\ge$~100~PeV) are expected to be produced 
from the decay of charged pions created in the
interactions between ultra-high energy cosmic rays and 
cosmic microwave background photons~\cite{bz}.
Both populations of high-energy neutrinos combine to offer a unique probe,
spanning many orders of magnitude in energy, of the highest energy astrophysical phenomena in the universe.

High-energy neutrinos 
can be detected in the VHF-UHF radio bands ($\sim$10-1000~MHz) 
through the highly impulsive radiation generated by neutrino-induced electromagnetic showers in dense dielectric media.  
This coherent radio emission is caused by the Askaryan effect,
whereby a $\sim$20\% negative charge excess develops, through positron annihilation and
other electromagnetic scattering processes, as the shower traverses the media faster
than the local light speed~\cite{askaryan1, askaryan2, zhs}.
The Askaryan effect has been confirmed in a series
of beam tests using sand, salt, and ice as target materials~\cite{sand, salt, ice}. 
Glacial ice is a good detection
medium because of its $\sim$1~km attenuation length at
radio frequencies smaller than 1~GHz~\cite{spradioatten,mooresbayradioatten,summitradioatten}. 

The Antarctic ice sheet provides the necessarily large volumes for
radio detectors in search of neutrino-induced Askaryan emission~\cite{antforneut}.
The ANITA experiment is a long-duration balloon
payload with high-gain antennas, which instruments $\sim$100,000~km$^{3}$ of
ice while circumnavigating the Antarctic continent at
float altitude and has an energy threshold of $\sim$10$^{3}$ PeV~\cite{anitaInstrument}.
The ground-based experiments of ARA and ARIANNA, 
both in early stages of development, are composed of a number of
independent radio-array stations that will reach energies down to 50-100~PeV at
full design sensitivity~\cite{araInstrument, ariannaInstrument}.
Installing
antennas as close as possible to the neutrino interaction is key to increasing the sensitivity
at lower neutrino energy.
At present, the ANITA experiment provides the best limits for diffuse fluxes
of high-energy neutrinos 
with energies above $\sim$10$^{5}$~PeV~\cite{anita18},
while IceCube sets the best limits 
at lower energies down to their detected flux, around 1~PeV~\cite{IceCubeHES}.

The radio detection method offers a way forward to the $\ge$10~gigaton scale
detectors required to detect and study high-energy neutrinos 
at energies beyond the flux measured by 
IceCube due to the much longer attenuation and scattering 
lengths at radio compared to optical wavelengths.
Ground-based radio detector stations can be separated by as 
much as a few kilometers, with each
station monitoring an independent volume of ice so that the total 
active detection volume scales linearly with the number of stations.

In general, for a given number of receivers with specified bandwidth,
the lowest energy threshold will be achieved by:
\begin{enumerate}
 \item Placing radio receivers as close as possible to the neutrino interaction
 \item Using a medium with a long radio attenuation length, such as South Polar glacial ice
 \item Maximizing directional gain while preserving a
   wide overall field of view (for a diffuse neutrino search).
\end{enumerate}

A radio detector with improved low energy sensitivity will dig into the
falling spectrum of astrophysical neutrinos
observed by IceCube~\cite{IceCubeAstrophysical}. 
These astrophysical neutrinos, 
as opposed to the cosmologically produced neutrino population,
are unique messengers in the realm of multi-messenger astrophysics due to 
being created promptly in and traveling unimpeded from 
the highest energy particle accelerators in the universe.
Additionally, reaching the 10~PeV threshold would provide meaningful
energy overlap with the IceCube detector, which would provide 
insightful cross-calibration of the
radio detection technique with established optical 
Cherenkov high-energy neutrino detectors~\cite{phasedarrayconcept}.

\section{Radio Array Triggering}
\label{subsec:triggering_intro}
In order to reconstruct the energy and direction of the 
high-energy neutrino from its radio-frequency (RF) emission, 
it is important to precisely measure the relative timing, 
polarization, and amplitudes received at an antenna array.
Ultimately, to best extract these low-level observables, 
it is necessary to save the full Nyquist-sampled waveforms,
which requires several gigasamples-per-second (GSa/s) recording of each antenna output. 
It is not possible to continuously stream data to disk at these rates, 
so events must be triggered.

The signature of neutrino-induced Askaryan emission is a 
broadband impulsive RF signal,
which is inherently unresolvable over the entire spectrum of emission in
a realistic radio detector system~\cite{zhs, ice}.
The received signal is band-limited such that the characteristic pulse time resolution, $\Delta{t_{BL}}$,
is approximately equal to 1/(2$\Delta{\nu}$), where $\Delta{\nu}$ is the receiver bandwidth.
For an in-ice receiver, band-limited thermal
noise is also measured from the ice ($\sim$250~K) 
and introduced by the system ($<$~100~K, typically). 
Therefore, the detector-level sensitivity
is determined by the efficiency at which the trigger 
system is able to accept Askaryan impulses
over fluctuations of the thermal noise background.

The ANITA, ARA, and ARIANNA detectors have approached 
triggering with a fundamentally similar
strategy~\cite{anitaInstrument, araInstrument, ariannaInstrument}.
The signal from each antenna, either in voltage or converted to power,
is discriminated on the basis of a single or multi-threshold level to form an antenna-level trigger.
A global station (or payload, in the case of ANITA) trigger is formed
using a combinatoric decision based on a minimum number of antenna-level (or intermediate)
triggers in a causal time window determined by the geometry of the antenna array.
This method has low implementation
overhead as it only requires a per-antenna square-law detector and a single
field-programmable gate array (FPGA) chip to perform the thresholding 
and trigger logic~\cite{varnerfpga, SST}.

This triggering scheme performs well in rejecting accidentals 
caused by random thermal noise up-fluctuations;
these systems can typically trigger efficiently, 
while keeping high detector live-time,
on 3-4~$\sigma$ radio impulses, where~$\sigma$~is the RMS level of the thermal noise
background~\cite{anitaInstrument, araInstrument, ariannaInstrument}.
However, the trigger is essentially limited to 
coherently received power in effective apertures
defined by a single antenna element in the detector array\footnote{
  The effective aperture, $A_{eff}$, of an antenna is given by $ G\lambda^2 / 4 \pi$,
  where $\lambda$ is the wavelength and $G$ is the directive antenna gain in linear units.}.

\subsection{An interferometric trigger}
\label{subsec:newtrigger}
A coherent receiver with a larger aperture can be made by
using a single equally high-gain antenna, or by
the interferometric combination of signals from lower-gain antennas. 
The latter technique of
aperture synthesis is widely used in radio astronomy for increasing
angular resolution
of a telescope beyond what is feasible with a single high-gain dish antenna.
In many cases, the interferometric radio array is electronically steered using either 
time- or frequency-domain beamformers, also known as `phasing'.

\begin{figure}[]
\centering
\includegraphics[trim=4.5cm 2cm 7cm 1.5cm,height=9cm, clip=true]{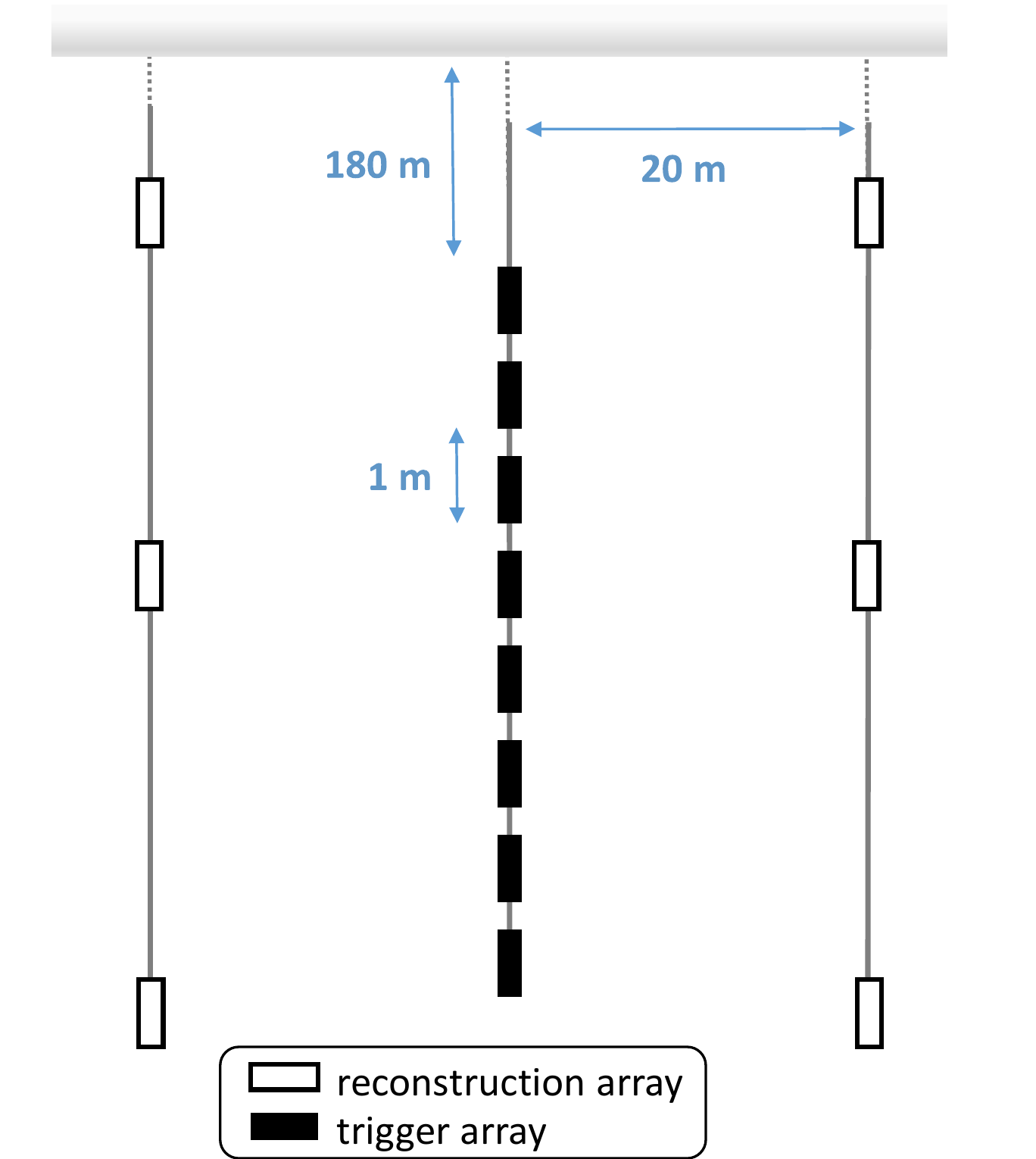}
\caption[conceptual side view]{Conceptual drawing of an in-ice radio array.
  A dedicated compact antenna array is used for an interferometric trigger system.
  Sparsely instrumented strings are placed at longer baseline spacings and used for improved
  angular resolution.}
\label{fig:concept}
\end{figure}

In the context of radio detection of high-energy neutrinos, we consider an in-ice
interferometric trigger system shown in Fig.~\ref{fig:concept} and proposed in~\cite{phasedarrayconcept}.
Geometric constraints of the drilled ice-borehole (diameter of $\sim$15~cm) 
typically limit the deployment to only low-gain antennas. It is 
possible to increase the effective gain at the trigger level by phasing, in real time,
the transduced voltages in the compact trigger array shown in Fig~\ref{fig:concept}. 

The array factor, $AF$, of a vertical uniform array composed of $N$ elements with spacing $d$,
impinged upon by a monochromatic plane wave with wavelength $\lambda$ and zenith angle, 
$\theta$, is given by
\begin{equation}
  \label{eqn:arrayfactor}
  AF(\psi) = \frac{\sin(\frac{N \psi}{2})}{\sin(\frac{\psi}{2})} ,
\end{equation}
where $\psi = 2 \pi d \cos{\theta} / \lambda$ as derived in~\cite{balanis}.
This factor describes the array directivity, given by $D_{array}(\theta) = |AF(\theta)| \:   D_{element}(\theta)$,
where $D_{element}$ is the directivity of the individual antennas and assuming a uniform azimuthal response.
For an array with element spacing of $\lambda/2$, $AF$ reaches a maximum of $N$ at
broadside ($\theta$ = $\pi/2$), such that the maximum array gain, in dBi, is 
\begin{equation}
  \label{eq:arraygain}
  G_{array} = 10 \: \log_{10}(N \: D_{element}(\pi/2)) .
\end{equation}
For a uniform array of 8 dipole antennas ($D$ = 1.64) with $\lambda$/2 spacing, the
maximum array gain is $\sim$11~dBi, comparable to the boresight gain of 
the high-gain horn antenna used on the ANITA payload~\cite{anitaInstrument}.
In general, the effective array gain will be frequency dependent because of 
the broadband nature of the RF~signal emitted by neutrino-induced showers. 

Time-domain beamforming methods are more suitable for wideband signals
and are used widely in ultra-wideband remote sensing, imaging, and
impulsive radar~\cite{uwbchipbeamformer2, timedarraysnutshell}.
The most common technique is delay-and-sum beamforming, which is described
by a coherent sum, $S(t)$, over an array of $N$ antennas as
\begin{equation}
  \label{eq:das}
  S(t) =   \sum_{n=0}^{N-1} \:w_{n} \: y_{n}(t-\delta_{n}) ,
\end{equation}
where $w_{n}$ is the weight applied to the antenna amplitude,
$y_{n}$ is the timestream signal of the antenna, and $\delta_{n}$ is the applied delay.
We use equal antenna weights for the beamforming trigger system described here,
such that the amplitude of correlated signals scales as $N$ in 
the correctly-pointed coherent sum,
while the uncorrelated thermal noise background only adds as $\sqrt{N}$.

Delay-and-sum beamforming can be implemented using switchable 
delay-lines or by the real-time processing of digitally-converted data. 
The delay-line implementation is optimal in terms of design cost
and power consumption in applications where only single beams are formed at any instant~\cite{uwbchipbeamformer2, mwa}.
However, delay-line methods become overly complex for applications
where multiple instantaneous beams are required, particularly
for those with $N_{beams} > N_{antennas}$; digital methods 
are preferred in these cases.

For a linear and uniformly-spaced vertical array,
the digital method can form full-array (using all antennas) 
coherent sums for received plane-wave
elevation angles, $\theta_{m}$, given by
\begin{equation}
  \label{eq:digitalbeam}
  \sin(\theta_{m}) = \frac{c\:m\:\Delta{t}}{n \: d} 
\end{equation}
where $c$ is the speed of light, $d$ is the element spacing,
$n$ is the index of refraction in the medium,
$\Delta{t}$ is the sampling interval of the digital data, and
$m$ is an integer, later referred to in this paper as the `beam number'. 


The beamwidth of a monochromatic receiving array of uniform 
element spacing $d$ is approximately given by
$\lambda / (Nd)$. To convert to wideband signals, 
a bandwidth $\Delta{\lambda}$ is considered and is
substituted for the characteristic band-limited timing resolution, $\Delta{t_{BL}} = 1/(2\Delta{\nu})$,
giving a beamwidth of
\begin{equation}
  \label{eq:beamwidth}
  \Theta_{FWHM} \simeq \frac{2 \: c \: \Delta{t_{BL}}}{n \: N\:d} 
\end{equation}
where $c$ is the speed of light, $n$ is the index of refraction,
$N$ is the number of baselines involved in the coherent sum, and $d$ is the uniform antenna spacing.

\subsection{The NuPhase Detector and Trigger}

Here we describe the design, implementation, and performance of
an interferometric trigger system
for radio-detection of high-energy neutrinos, which we call {\it NuPhase}. 
The trigger system consists of a linear array of low-gain antennas deployed
sub-surface in glacial ice, as depicted in Fig.~\ref{fig:concept},
whose signals are converted by low-resolution streaming ADCs
and fed into an FPGA for digital beamforming via coherent sums. 
The power in each beam is continuously measured in short 
$\sim$10~ns intervals in search of impulsive broadband radio signals,
generating a trigger signal for a separate reconstruction array provided by an ARA station.
The first complete detector was installed at the South Pole during the 2017-18 Austral summer season.
The NuPhase detector builds upon preliminary testing and simulation studies
reported in~\cite{phasedarraynim,abbyPA}.

In Sec.~\ref{sec:ARA}, we describe the installation of the NuPhase detector
as part of the ARA experiment at the South Pole.
The details of the NuPhase detector system, from the in-ice RF receivers to the
data processing, are given in Sec.~\ref{sec:detector}.
Sec.~\ref{sec:beamformingfirm} covers the beamforming strategy and the
firmware deployed on the processing FPGA. The performance of the beamforming
trigger is provided in Sec.~\ref{sec:trigger}.
In Sec.~\ref{sec:arasim}, we incorporate the measured performance in to
neutrino simulation studies to determine the achieved improvement in sensitivity. 
Finally, we conclude in Sec.~\ref{sec:conclude}.

\section{Installation with an ARA Station at the South Pole}
\label{sec:ARA}
The ARA experiment has at present 5 deep antenna stations~\cite{araInstrument,aratwostation}.
The baseline ARA station includes four instrument strings, each holding four antennas:
two horizontally polarized (Hpol)~+~vertically polarized (Vpol) antenna pairs. 
The antenna-pair vertical spacing on a single string is 20-30~m and the string-to-string spacing is 30-40~m, 
with the four strings installed in a rectangular pattern. 
Every station has at least one outrigger calibration pulser string that has both Hpol and Vpol transmitting antennas,
which can be fed by either a fast impulse or a calibrated noise source~\cite{araInstrument}.

\begin{figure}[]
\centering
\includegraphics[height=9cm]{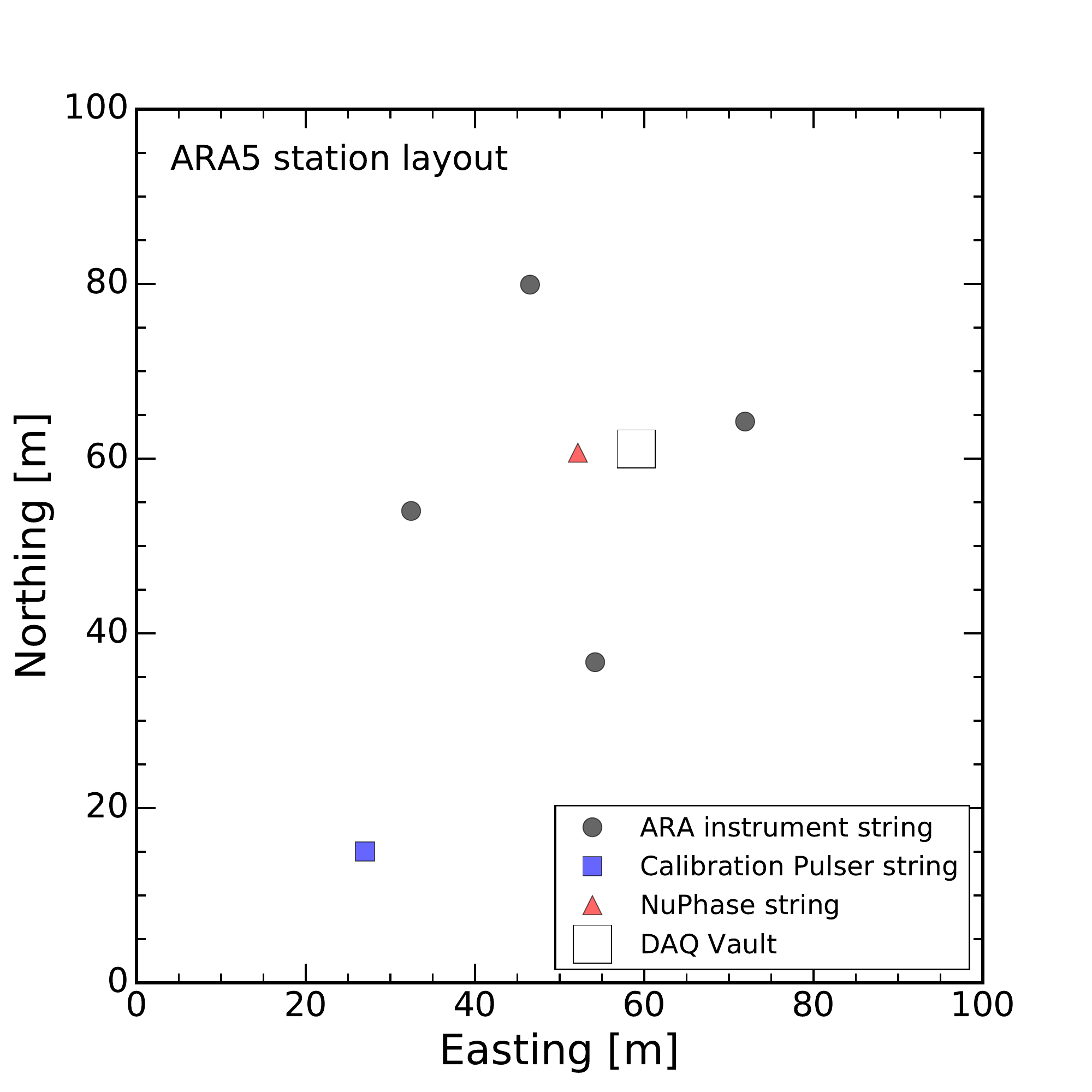}
\caption[A5 station layout]{
	ARA5 station layout of deep antenna strings. 
	Installed during the 2017/18 Austral summer season, the ARA5 station
    includes the NuPhase trigger string at the center of the station.
    The 4-antenna instrument strings have a closest baseline spacing of $\sim$40~m. 
    All detector strings at the station are deployed to a depth of 190-180~m below the surface.}
\label{fig:ARA5}
\end{figure}

The ARA signal chain splits into a trigger and signal path after full amplification.
The trigger path is sent through a tunnel diode, implemented as a square-law detector,
and the output integrated with a time constant of $\mathcal{O}(10~$ns$)$ 
and compared directly to an analog threshold at a differential FPGA input. 
In standard operation, the ARA trigger requires at least 3 out of 8 of either the 
Hpol or Vpol tunnel diode outputs above threshold within a few~100~ns window 
(depending on the specific station geometry).

The NuPhase antenna array is deployed at the center of the ARA5 station, as shown
in Fig.~\ref{fig:ARA5}. In this context, the NuPhase array serves as the `trigger'
array and the ARA array, with its much larger antenna baselines, serves as the 
reconstruction, or `pointing', array.  
The trigger output from the NuPhase electronics is 
plugged into the external trigger input of the ARA data acquisition (DAQ) system. 
Because the NuPhase detector generates only a Vpol trigger, ARA5
is configured to trigger on the logical OR of the NuPhase trigger and the standard
ARA trigger. The NuPhase and ARA5 DAQ systems
run on separate clocks. A GPS receiver at the ARA5 site 
synchronizes timing on second scales between the two instruments. 
Longer term timing ($>$second) relies on Network Time Protocol (NTP) 
synchronization of the two systems.  

\section{Detector Systems}
\label{sec:detector}

The NuPhase detector consists of several subsystems: 
the antenna array and RF signal chain,
the analog-to-digital conversion (ADC) boards, 
the FPGA firmware, power distribution,
and the acquisition software.

\subsection{RF Signal Chain}
\label{subsec:signal_chain}

\begin{figure*}[]
\centering
\includegraphics[trim=0.5cm 2.7cm 0.5cm 3.5cm, height=6.6cm, clip=true]{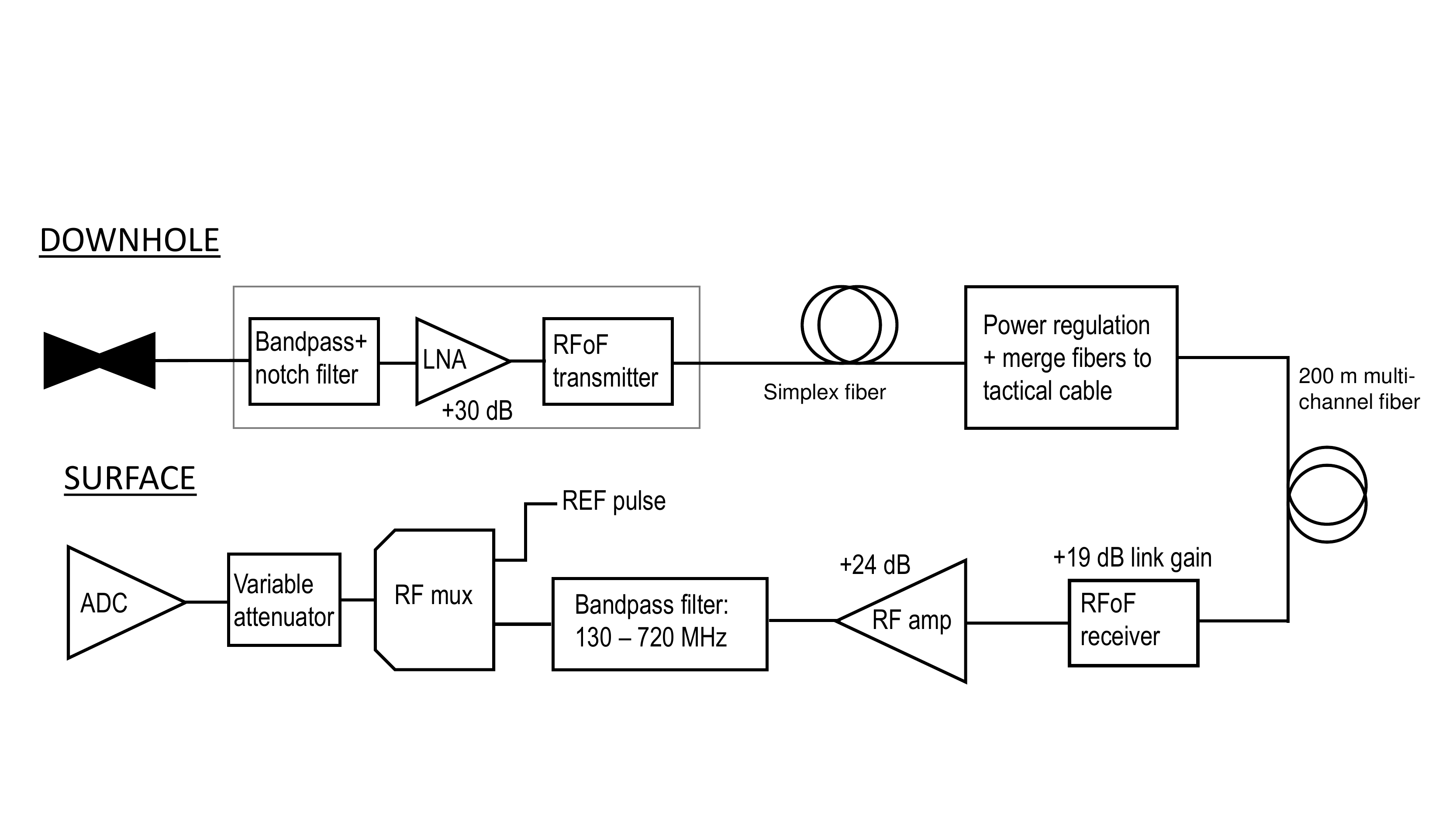} 

\caption[Diagram of the RF signal chain]{
	Single channel RF signal chain. A front-end amplifier module, which contains
  	a bandpass and 450~MHz notch filter, a low-noise amplifier (LNA),
  	and Radio-frequency-over-Fiber (RFoF) transmitter, is
  	co-located with each antenna (shown in Fig.~\ref{fig:lna_and_vpol}).
  	RF signals from each antenna are sent up through the array on a 
    MIL-SPEC single-mode fiber.
  	At the top of the array is a load-bearing cylinder that 
    holds the power regulation board and merges the individual optical 
    fibers to a tactical fiber bundle that sends the signals to the surface.
  	At the surface the signals are converted back to copper and
    sent through a last stage of amplification and filtering.
    The RF mux (ADG918) allows a fast FPGA-generated pulse (REF pulse) to be 
    inserted into the ADCs for timebase calibration and the digital attenuator
    allows for channel-to-channel gain balancing.
    Finally, the RF signal is inserted into the ADC/DAQ system diagrammed
    in Fig.~\ref{fig:nuphase_electronics_diagram}.}
\label{fig:signal_chain}
\end{figure*}

The full NuPhase RF signal chain is shown in Fig.~\ref{fig:signal_chain}.
A description of the signal chain, from
the antennas through the last stage filtering and amplification, follows.
  
The NuPhase antenna array is deployed down a single 200~m deep, 16~cm wide,
ice borehole located in the center of the ARA5 station.
A total of 12 antennas are installed: 
10 Vpol birdcage-style antennas along with 2~Hpol ferrite loaded quad-slot antennas. 
The Hpol antennas are identical to those used for ARA instrument strings
while the Vpol antennas have a different feed point design --
the ARA antennas are described further in~\cite{araInstrument}.
Both antenna types have approximately uniform azimuthal beam patterns.
The two Hpol antennas are deployed at the bottom of the NuPhase string with a spacing of 2~m,
followed by the 10~Vpol antennas at 1~m spacing.

The Vpol receiving antennas used in the beamforming trigger are 
relatively broadband, with good receiving sensitivity
in the range of 150-800~MHz and have a single-mode beam pattern below $\sim$500~MHz. 
The Hpol receiving antennas are not involved in the beamforming trigger, 
but are recorded to get a complete picture of the field polarization for each event.

\begin{figure*}[]
\centering
\subfloat[]{\includegraphics[height=3.8cm]{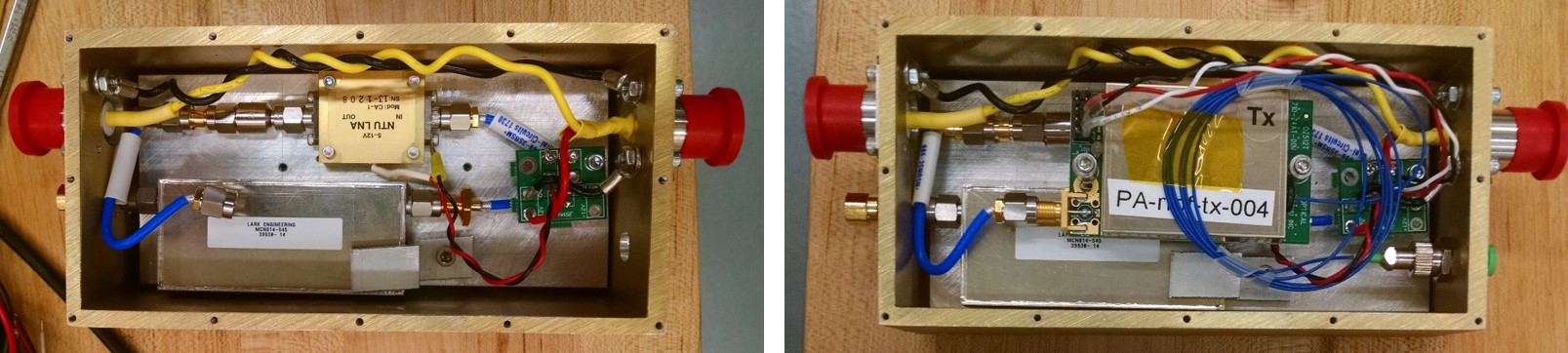}} \\
\subfloat[]{\includegraphics[height=3.4cm]{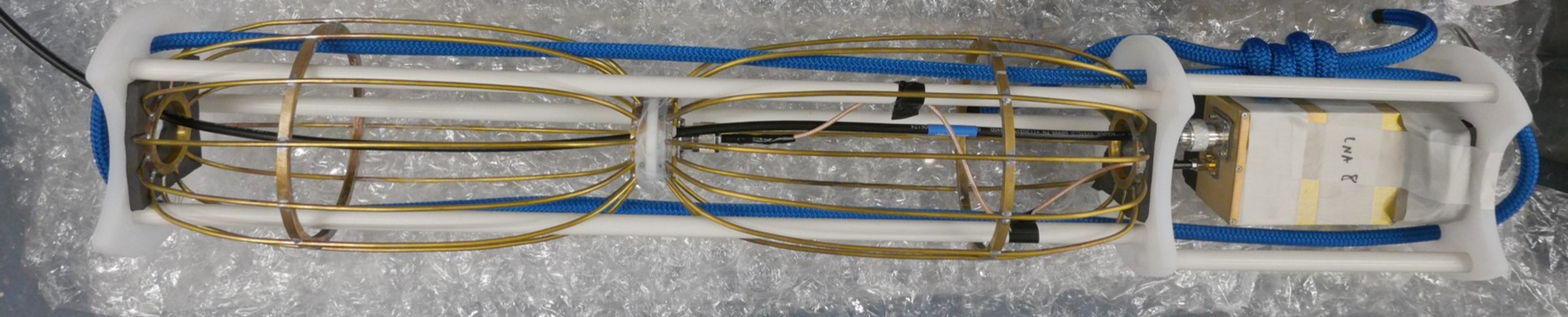}}  
\caption[]{(a) Front-end amplifier module. The photo on the left shows the integrated bandpass and notch filter, 
  the LNA, and the power pick-off board with in-line ferrites. 
  Each front-end amplifier receives power from the overhead 
  antenna unit and passes through to the unit below. 
  On the right, the same box is shown with the RFoF transmitter installed and 
  connected to the fiber feed-through adapter.
  (b) A fully assembled vertically-polarized (Vpol) antenna unit.
  The birdcage antenna and front-end amplifier are installed in a 
  frame constructed of fiberglass rods and 
  ultra-high molecular weight plastic faceplates.
  The frame measures 88~cm in length with a diameter of 15~cm. A length of coaxial
  cable, used for DC power, is routed through the antenna feed and connected to the
  power pass-through input/output of the front-end amplifier.}
\label{fig:lna_and_vpol}
\end{figure*}

A front-end amplification module, including a 
low-noise amplifier (LNA), bandpass filter,
and RF-over-fiber (RFoF) transmitter,
is embedded with each antenna, comprising an `antenna unit'. 
The front-end amplification module and the
assembled Vpol antenna unit are shown in Fig.~\ref{fig:lna_and_vpol}.  
The compact design allows the antenna units to be deployed 
at a spacing of one meter. During deployment, each antenna unit
required only two connections: 
the N-type coaxial cable power connection and a single mode
optical fiber carrying the RF signal.
The RFoF system is required to send high-fidelity 
broadband signals over the 200~m distance
from the antennas in the ice boreholes to the electronics at the surface.

The LNA provides 32~dB of gain with an intrinsic noise 
figure of $\le$~0.6~dB over the band.
In combination with the short antenna feed cable ($\sim$0.3~dB) and the
relatively noisy RFoF link system ($\sim$25~dB),
the noise figure increases to roughly 1.4~dB 
over the system bandwidth.
A band-defining filter is placed before the LNA,
which has extremely low insertion loss except for a deep $\sim$50~dB notch
at 450~MHz to suppress land mobile radio communications 
used widely around South Pole Station.

The front-end amplifier module also serves as a pass-through for the array power,
which simplifies the wiring during deployment and, crucially, 
ensures that the complex impedances of the Vpol antennas in the array are matched. 
Each Vpol antenna in the array
has a single Times Microwave LMR-240 coaxial cable running 
through the antenna feed that passes power to the next antenna unit. 
Signal outputs from the amplifier modules are routed up through higher antennas
on optical fiber, which have negligible influence on the antenna response.
We ensure that the impulse responses of the antennas will be the same by matching 
the internal metal wiring, thus optimizing a beamforming trigger. 

Each front-end amplification module draws 200~mA on a 12~V supply, 
dominated by the RFoF transmitter, so
that the total power draw of the NuPhase downhole array is roughly 25~W.
At the top of the array is a custom power regulation board 
designed to operate down to -55$^{\circ}$C, 
which is housed in a load-bearing RF-shielded cylinder. 
The downhole power board
linearly regulates to 12.5~V (allowing for IR losses along the array),
sourced by an efficient switching power supply at the ice surface. 
Transient switching noise is suppressed both by filtering circuits 
on the downhole power board and parasitic resistance and inductance on
the long 200~m coaxial cable that transmits from the surface.
We find no evidence of power transient induced triggers in our system.

The RF signals are sent to the surface over a 200~m long 12-channel tactical fiber.
A bank of RFoF receivers are installed in the NuPhase instrument box at the surface,
which convert the signals back to standard copper coaxial cable.
A custom second-stage amplification and filtering board supplies the 
last 20~dB of signal gain, while
filtering out-of-band LNA noise and ensuring at 
least 10~dB of anti-aliasing suppression at 750~MHz.
Lastly, a digitally-variable attenuator is
placed on each channel that is used to match
overall gains between channels and to tune the digitization resolution.

\begin{figure}[]
\centering
\includegraphics[trim=0cm 1.2cm 0.8cm 1cm, height=11cm, clip=true]{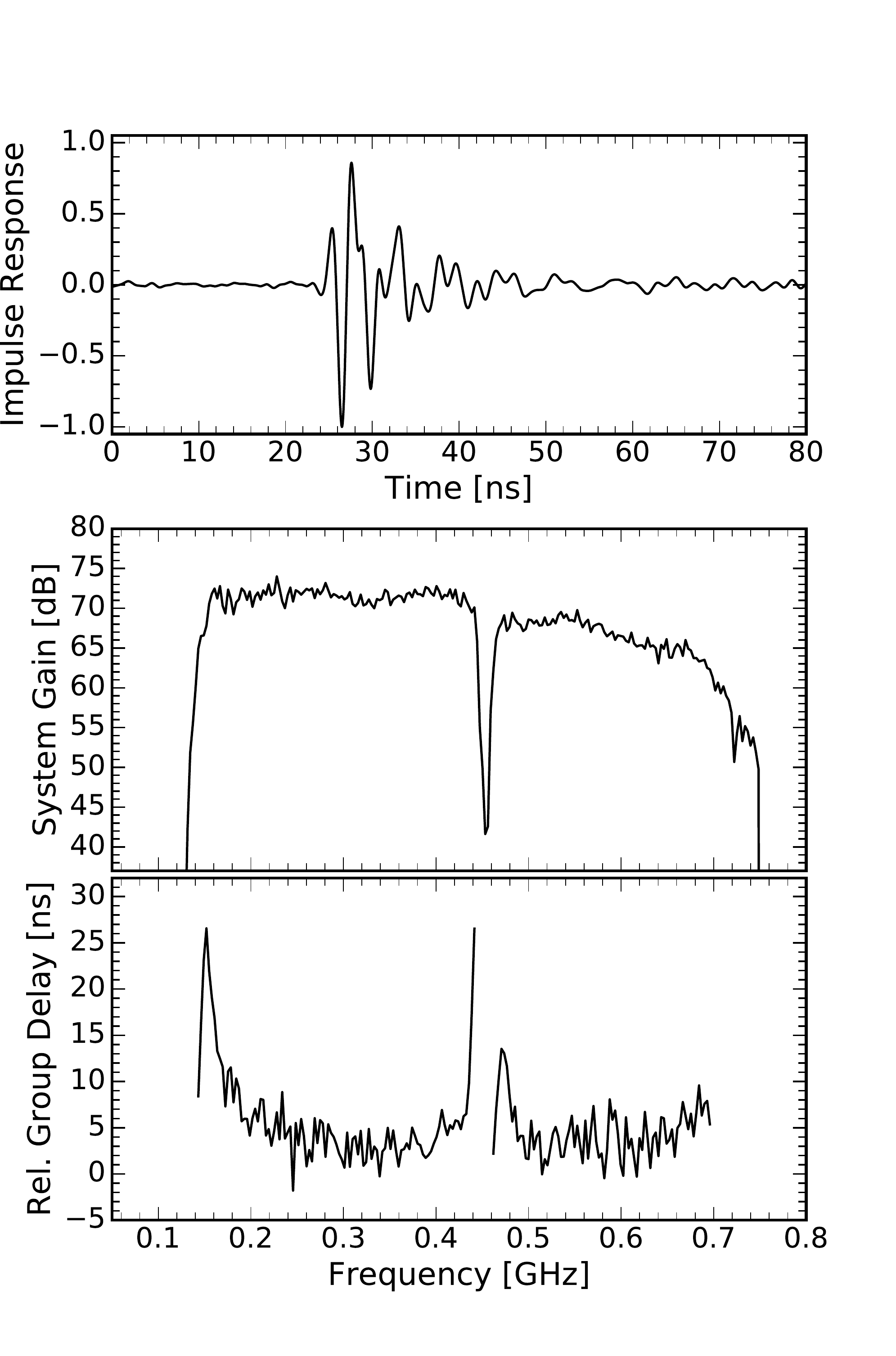}
\caption[]{The NuPhase RF signal chain response.
		The top plot shows the time-domain response of the 
        full NuPhase signal chain, excluding the antenna.
        The gain magnitude is roughly flat at 70~dB
        between the 150~MHz low-edge and the 450~MHz notch filter, 
        and rolls off at higher frequencies
        due to the second-stage filter and the differential amplifier stage
        on the ADC board. The relative group delay is plotted at frequencies
        where the gain magnitude exceeds 10\% of the value between 200 and 300~MHz.}
\label{fig:sys_response}
\end{figure}

The full RF signal chain response was measured using a fast impulse
from an Avtech AVP-AV-1S pulse generator. Several thousand pulses were digitized
and recorded using the DAQ system described in the next section. 
The impulse response is found by deconvolving the 
Avtech input pulse from the recorded signal 
and is shown in Fig.~\ref{fig:sys_response}.
The system reaches a peak gain of $\sim$70~dB in the 150-450~MHz band
and rolls off to 64~dB at 700~MHz due to both the 
second-stage filter ($\sim$2~dB) and the active 
differential amplifier stage on the ADC board ($\sim$4~dB).
A wideband balun will be used in place of the 
differential amplifier for any future designs.
Impulse response dispersion is produced at the edges of the 
high-pass and 450~MHz notch filters.

\begin{figure*}[]
\centering
\includegraphics[trim=2cm 2.5cm 2cm 3.5cm, height=7.6cm, clip=true]{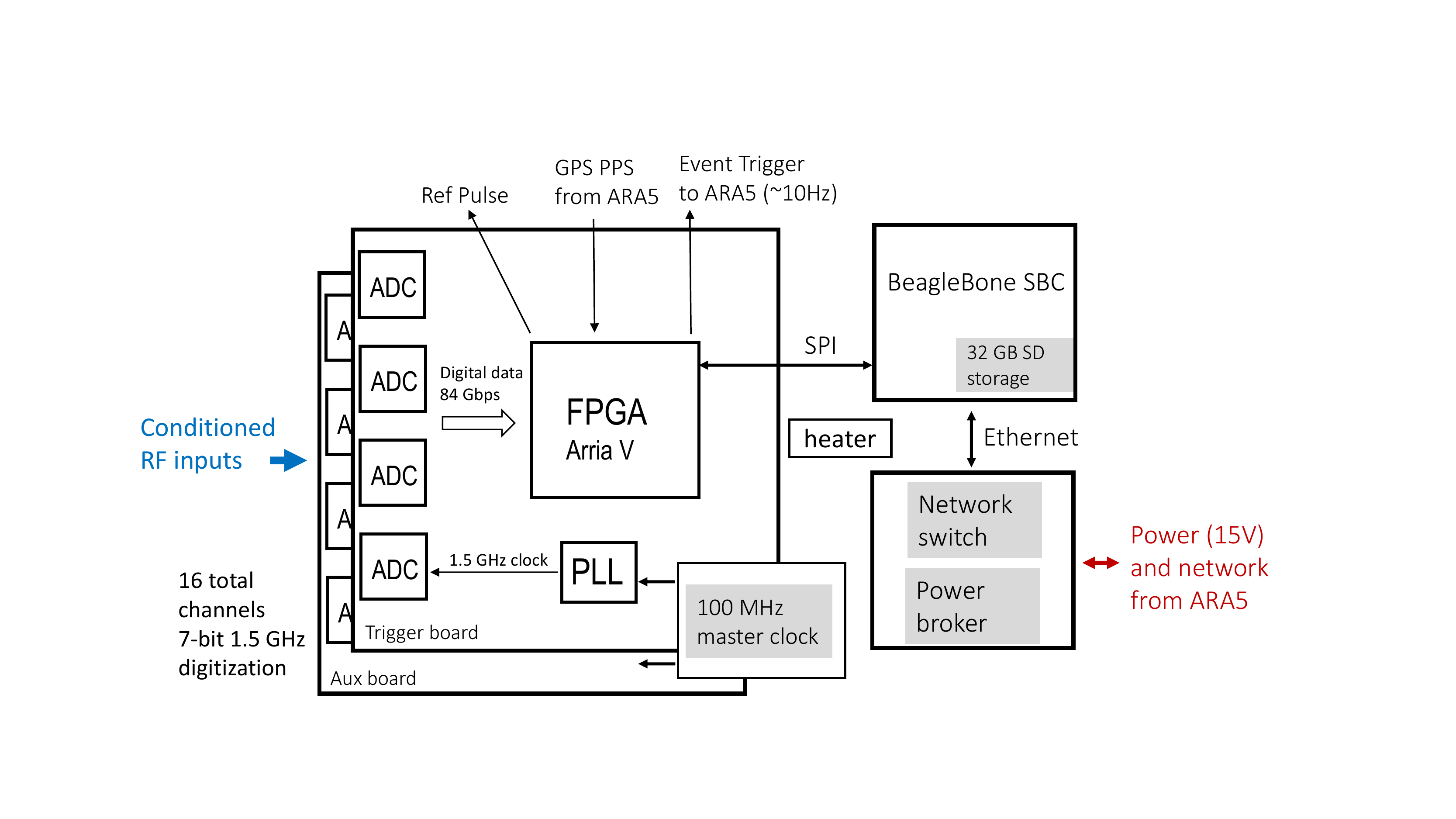} 
\caption[Diagram of nuphase digital electronics]{Overview of the NuPhase DAQ. 
	The system includes two ADC boards with a total of 16 channels 
    of 1.5 GSa/s digitization at 7-bit resolution.
  	Both boards save and transmit full waveforms to the single-board computer (SBC), 
    but only the trigger ADC board includes the full beamforming firmware.}
\label{fig:nuphase_electronics_diagram}
\end{figure*}

\subsection{Data Acquisition System}
\label{subsec:digital}

\begin{figure*}[]
  \centering
  \includegraphics[trim=2.5cm 1cm 2.5cm 1cm,height=11cm, clip=true]{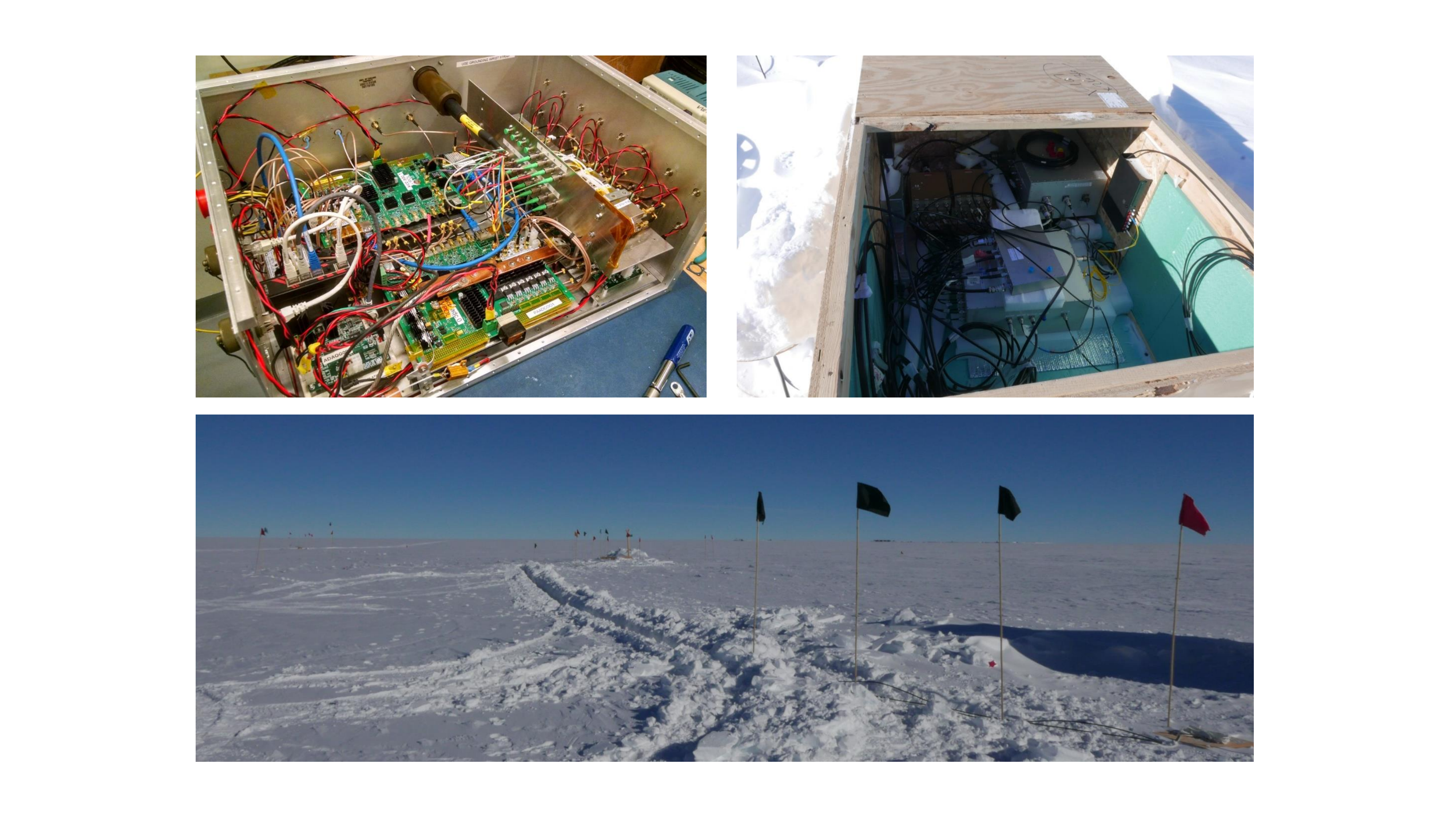} 
  \caption[NuPhase instrument box / SP photos]
          {Upper left: The NuPhase instrument box in the lab.
            Upper right: The ARA5 surface vault pictured shortly after installation.
            The vault houses the NuPhase instrument,
            the ARA5 DAQ box, the ARA 300~W power step-down converter box,
            and patch panels for both network fiber and power.
            Bottom photo: 
            The ARA5 station, as viewed on the surface from 
            the calibration pulser borehole,
            is located $\sim$6~km from South Pole Station, 
            visible in the distance in the center-right of the image.} 
\label{fig:surface_electronics}
\end{figure*}

The NuPhase DAQ and trigger system is housed in the same RF enclosure 
(the `instrument box') as the second-stage amplifier boards. 
An overview drawing of the NuPhase DAQ is shown in Fig.~\ref{fig:nuphase_electronics_diagram}.
The instrument box is shown in Fig.~\ref{fig:surface_electronics}.


\begin{figure*}[]
\centering
\includegraphics[trim=2cm 0.5cm 0cm 0cm, height=6.9cm]{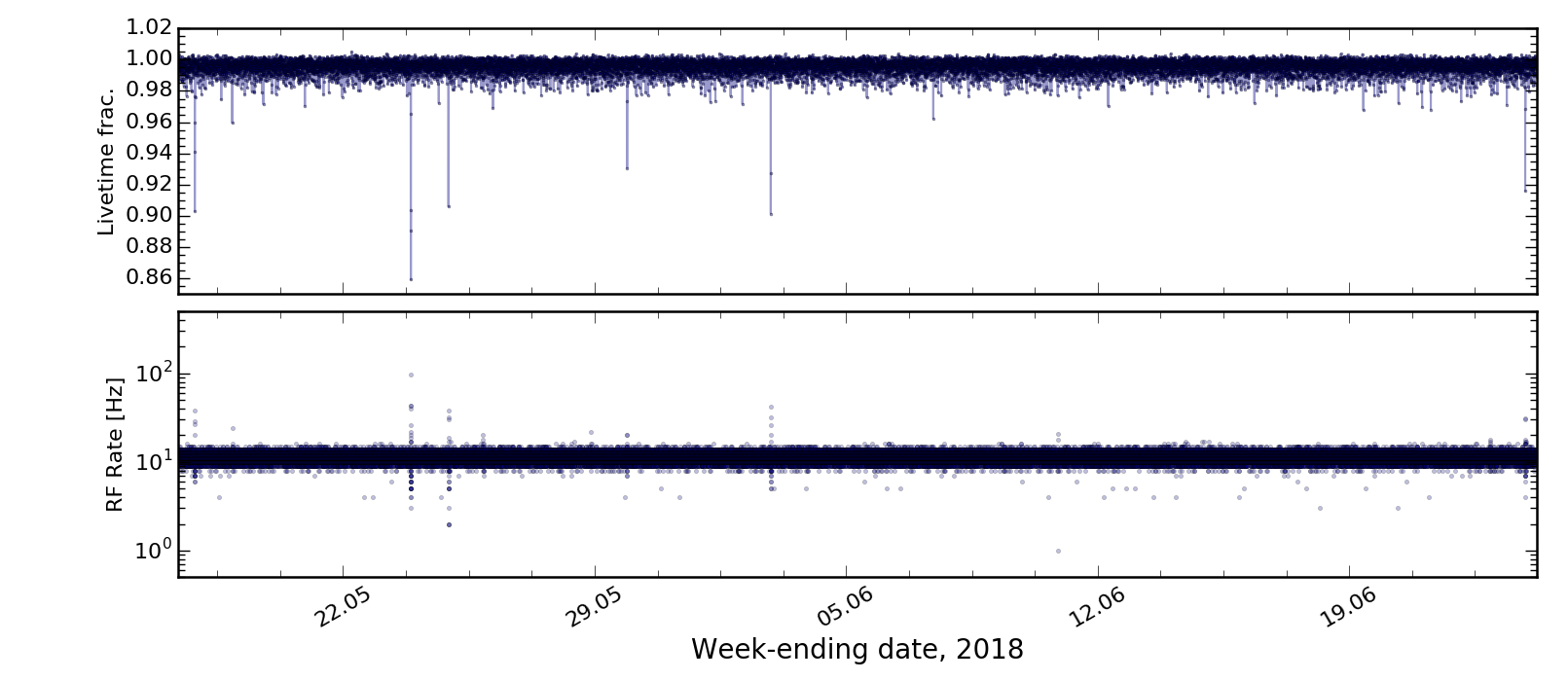} 
\caption[steady-state performance]{NuPhase RF trigger rate and livetime during
a month of operation in 2018. The occasional spikes in 
trigger rate are due to weather balloon launches at South Pole Station, when
we observe a correlated spectral line at 405 MHz, 
the carrier frequency of the balloon radiosonde.}
\label{fig:nuphase_rates}
\end{figure*}

A pair of 8-channel custom ADC boards serve as the workhorse of this detector. 
These boards use commercially available digitizers\footnote{Texas Instruments ADC07D1520} 
to convert data at 1.5 GSa/s with 7-bit vertical resolution. 
The ADC boards accept single-ended signals, which are
converted using a unity-gain differential amplifier stage.
The ADC output data streams are wired directly to LVDS receivers 
on a high-performance Intel Arria~V FPGA.
In order to synchronize the ADC boards, 
a separate board holds a 100~MHz oscillator that
serves as a master clock for the system. 
This clock is up-converted to 1.5~GHz locally on the ADC boards
using a PLL chip\footnote{Texas Instruments LMK4808}.
Both the trigger and auxiliary boards include the same baseline firmware for 
system management and data recording,
but only the trigger board is programmed
with the beamforming firmware. 
The NuPhase Vpol channels are inserted into the trigger board
and the Hpol signals are recorded in the auxiliary board.
The generated trigger signal on the interferometry board is sent to the ARA5 DAQ, 
which requires a trigger latency $\lessapprox$~700~ns. 
During nominal operation, we set the target trigger rate to 0.75~Hz in
each of the 15~beams, for a total RF event rate of $\sim$11~Hz.

It is necessary to time-align the datastreams between the ADC chips
because there is a random 1.5~GHz clock-cycle offset on power-up.
This is done by outputting a fast pulse using a 
double-data rate output driver on the FPGA,
which is sent through a series of splitters 
and injected into each channel through an RF~switch
as shown in Fig.~\ref{fig:signal_chain}.
An FPGA-alignment procedure is performed at the beginning of 
each new NuPhase run to ensure the beamforming delays are well defined.

The firmware also includes four separate event buffers, 
which allow simultaneous writing from the
FPGA to the single-board computer (SBC) and recording of events in the FPGA.
Due to the nature of the thermal noise background,
multi-event buffering is important to reduce system
deadtime from close-in-time noise up-fluctuations. 
As implemented, each event buffer can hold up to 2~$\mu$s of continuous waveform,
which in total uses about $\sim$10\% of the available memory resources in the FPGA. 

The FPGA communicates with a 
BeagleBone Black SBC\footnote{https://beagleboard.org/black}, 
which is rated for operation to -40$^{\circ}$C, 
over a four-wire SPI interface clocked at 20~MHz. 
The livetime, defined as the fraction of time in which there is at least 
a single available event buffer on the FPGA, is consistently above 98\% at a 
steady event rate of 10~Hz while recording 300~ns duration waveforms, 
as shown in Fig.~\ref{fig:nuphase_rates}. 
Rates up to 30~Hz were tested with a $\sim$20\% loss in livetime.
The SPI interface is also used for remote re-programming of the FPGA firmware.

Low-voltage power is provided in the ARA5 vault using a 
dedicated 300~W 15~V power box designed for the newer ARA stations, which steps down
the 400~VDC sent from a power supply in the IceCube laboratory. 
The NuPhase instrument box draws $\sim$80~W at full operation running 10 out of the 
16 ADC channels. 

\begin{figure}[]
\centering
\includegraphics[trim=0cm 1cm 0cm 0cm, scale=0.42, clip=true ]{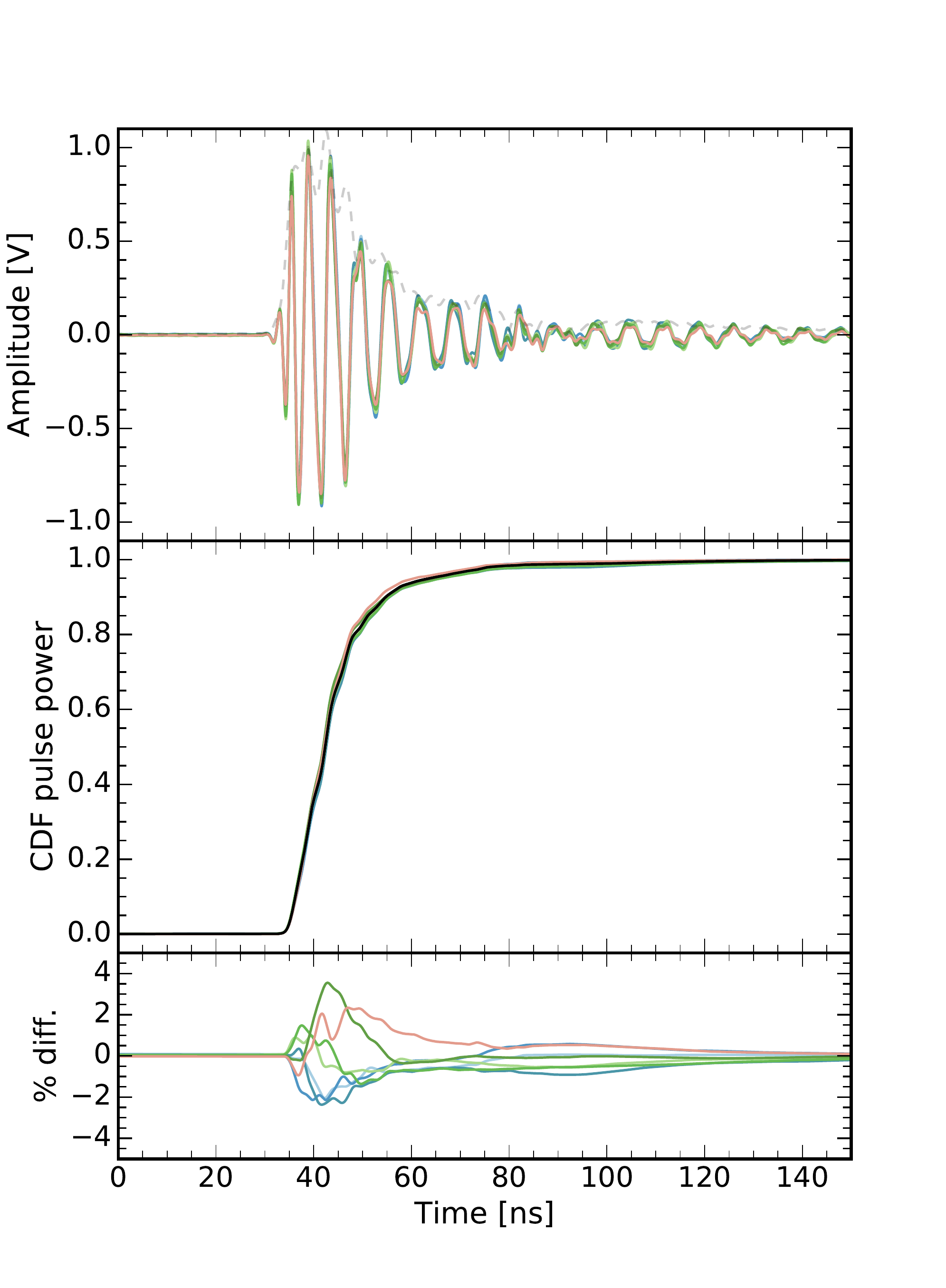}
\caption[cal pulser response]
	{Time-domain response to the {\it in situ} calibration pulser. 
	The top panel shows the measured calibration pulse on 
    each of the 7 Vpol 	channels.
  	Dispersion at the edges of the filter band (Fig.~\ref{fig:sys_response}) cause the 
    response to extend 	over $\sim$50~ns,
  	but 80\% of the power is held in the first 10~ns of the signal.  
    The middle panel shows the cumulative distribution
    function (CDF) of the normalized power in the pulse as a function of time. 
    The 7-channel average is shown by the black curve.
  	The channel-to-channel difference between the time-aligned
  	Hilbert envelopes and the 7-channel average 
    is sub-5\% as shown in the bottom panel.}
\label{fig:time_domain_response}
\end{figure}

\subsection{Acquisition Software}
\label{subsec:sw}

The BeagleBone Black SBC runs a Linux operating system (Debian 8.8) loaded from
a 32~GB SD card. The acquisition software is implemented in \texttt{C} as a set of
\texttt{systemd} units, allowing the use of standard built-in logging,
watchdog, and dependency facilities. The SBC has no persistent clock and is
reliant on NTP servers in the IceCube laboratory for time. 

The primary acquisition daemon (\texttt{nuphase-acq}) is responsible for
communicating with the FPGA over the SPI link. This multithreaded program uses
a dedicated thread with real-time priority to poll the board for available
events and read them out.  SPI transactions are performed in user-space, but
the \texttt{SPI\_IOC\_MESSAGE ioctl} provided by the \texttt{spidev}\footnote{https://www.kernel.org/doc/Documentation/spi/spidev} driver is
used pool up to 511 SPI messages per system call. 
While somewhat less efficient than
a dedicated kernel driver, working in user-space allows for rapid development
and increased robustness. Once read, events are buffered in memory using a
lockless circular buffer until a lower-priority thread can write events to the
SD card in a compressed binary format. 
Another lower-priority thread sends
software triggers and polls the board for beam-rate counters and adjusts
thresholds (Sec.~\ref{subsubsec:pid}) to maintain specified rates. 

Several additional programs complete the software stack. A `housekeeping
daemon' (\texttt{nuphase-hk}), records the state of the SBC and system currents
and temperatures. \texttt{nuphase-copy} is responsible for transferring files
to a server at IceCube Lab (from where data is sent North and archived to tape
using IceCube's JADE queue~\cite{jade}) and deleting already-transferred files.
The startup of the previously mentioned programs is contingent on the
successful execution of \texttt{nuphase-startup} which assures that board
temperatures are safe for operation of the FPGA and loads the latest 
application firmware.

\subsubsection{Trigger Rate Stabilization}
\label{subsubsec:pid}

Each trigger beam is assigned a rate goal (typically 0.5-1.5 Hz). 
To maintain the rate goal, the acquisition software uses a PID loop. 
Every second, the current trigger rate in each beam is estimated from a weighted average of
10-second counter values and a running average of 1-second counter values provided by the firmware. 
The `gated' trigger-rate counter (gated on the GPS second) is subtracted from the total to avoid
counting GPS-timed calibration pulser events in the threshold estimate. 
This estimate is compared to the goal, and the threshold for the beam is adjusted by a factor
proportional to the instantaneous error (P-term) and a factor proportional to
the accumulated error (I-term). The D-term is not currently used. 
The maximum increase in threshold is capped
in order to prevent a short burst of events from 
setting the threshold too high. 

\subsection{Calibration with Radio Pulsers}
\label{subsec:cal_pulser}
The ARA5 station is equipped with a calibration pulser string that includes
a fast pulse generator and a remotely-selectable Hpol or Vpol transmitting antenna.
The pulse width is $\sim$600~ps, as measured at the connection between the cable
and the transmitting antenna feed, providing a broadband calibration signal for 
the receiving array. The calibration pulser is installed at a depth of $\sim$174~m
and is located at a horizontal distance of $\sim$55~m from the NuPhase antenna array.

Fig.~\ref{fig:time_domain_response} shows the averaged waveforms recorded in 
each NuPhase Vpol channel using the ARA5 calibration pulser. 
The bulk of the signal power is contained in the first 10~ns of the waveform.
The channel-to-channel variation in the response is below the 5\% level,
which is important for the coherent summing trigger.

\begin{figure}[]
\centering
\subfloat[]{
\includegraphics[trim=10.5cm 4.7cm 10cm 6.6cm,height=4.4cm, clip=true]{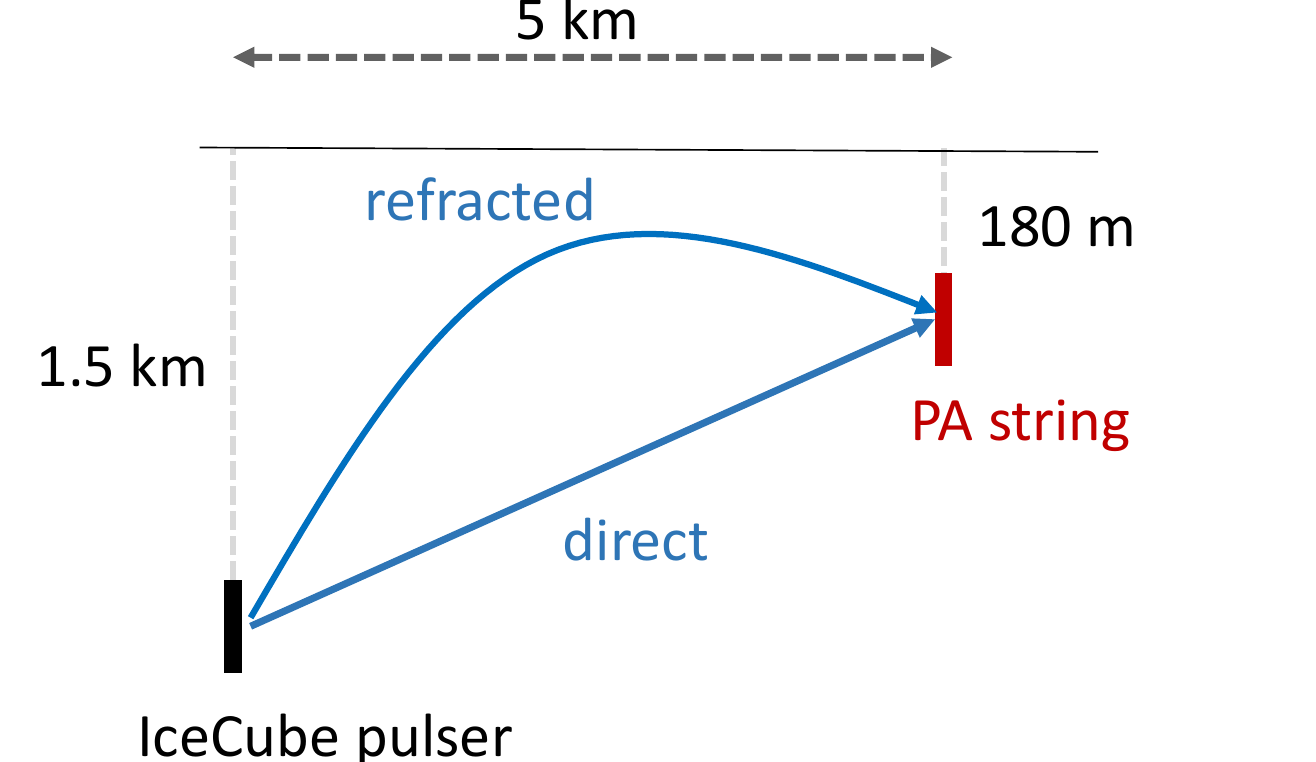}} \\
\subfloat[]{
\includegraphics[trim=0.7cm 0.76cm 0cm 2.50cm,height=9.2cm, clip=true]{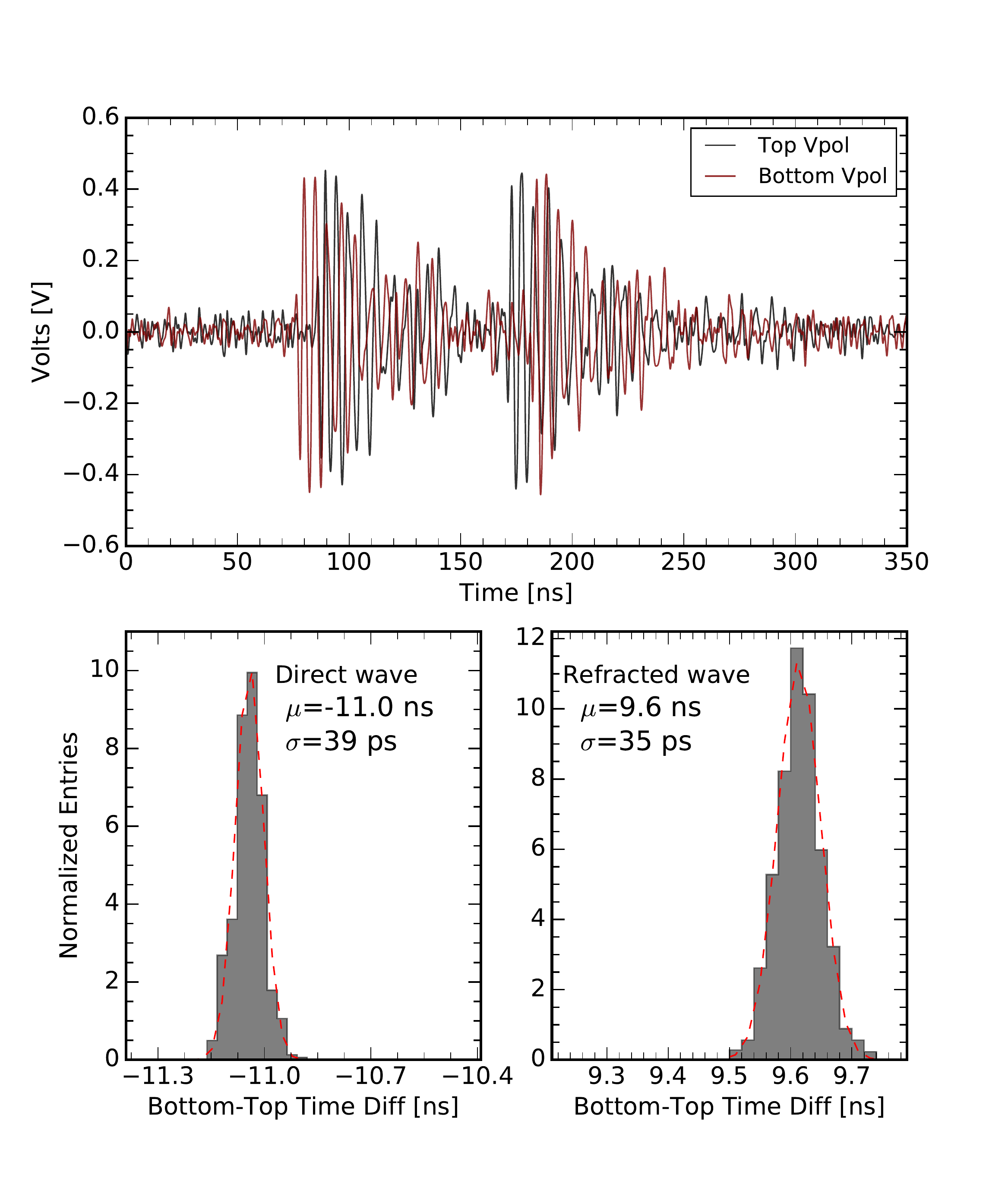}}
\caption[Deep Pulser]{
	a) The IceCube deep radio pulser, showing the direct and 
    refracted paths.
    b) Received pulses at the top and bottom Vpol antennas in the NuPhase array. 
    The histograms show the bottom-top Vpol time-difference for both 
    the direct and refracted radio pulses. 
    The two-channel system timing resolution is $<$40~ps. The direct plane-wave impulse
    was used to correct for timing mismatches shown in Fig.~\ref{fig:timing}.}
\label{fig:deeppulser}
\end{figure}

\begin{figure}[]
\centering
\includegraphics[trim=0cm 0.5cm 0cm 0.5cm,scale=.45,clip=True]{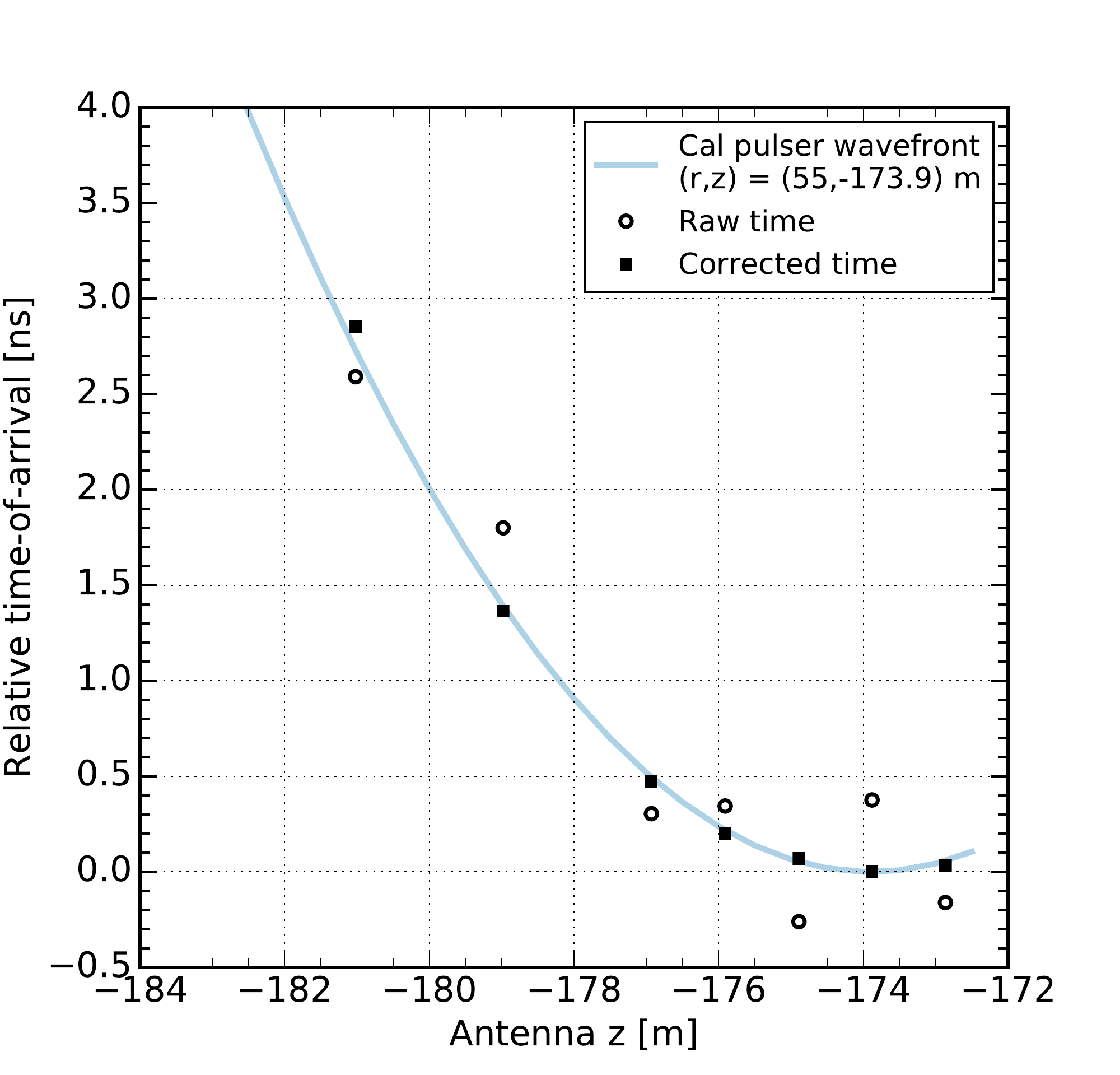}
\caption[cal pulser wavefront timing]{
	Reconstructed timing of calibration pulser events.
  There is a non-negligible systematic timing mismatch between channels and
  the measured time is shown pre- and post-timing correction.
  The timing correction is extracted using
  a separate dataset of far-field planewave events from the 
  IceCube deep radio pulser (Fig~\ref{fig:deeppulser}).
  The expected wavefront from a point source at the calibration pulser 
  location is overlaid.}
\label{fig:timing}
\end{figure}

A Vpol bicone transmitting antenna was installed at a depth of 
1450~m on IceCube string 22 during the construction of the IceCube detector
that, at a distance of $\sim$5~km from the ARA5 station, 
serves as an in-ice far-field calibration signal.
At this distance, the NuPhase array receives both a direct radio pulse
and a refracted (or reflected) pulse due to the index of refraction
gradient in the Antarctic firn, as shown in Fig.~\ref{fig:deeppulser}a.
An IceCube pulser event as recorded in the top and bottom Vpol antennas 
of the NuPhase array is shown in Fig.~\ref{fig:deeppulser}b, 
which clearly shows the direct and the refracted pulses.
The top and bottom Vpol antennas are separated by 8~m.
The bottom-top time difference shows the up-going and down-going
inclination of the direct and refracted pulses, respectively.
From several thousand pulser events, we measure the 
system time resolution by up-sampling the waveforms 
in the frequency domain and finding the peak in their discrete cross-correlation. 
The two-channel timing resolution on these high signal-to-noise ratio (SNR) pulses 
is found to be $<$40~ps.

\begin{figure*}[]
\centering
\includegraphics[scale=.55]{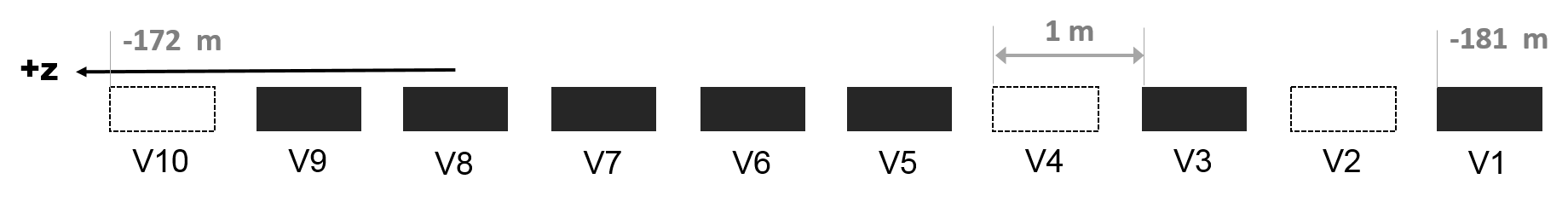} 
\caption[Vpol trigger array geometry]{
  Diagrammatic view of the Vpol trigger array. Ten Vpol antenna units, shown in Fig.~\ref{fig:lna_and_vpol}b,
  were deployed at 1~meter spacing starting at a depth of 181~m.
  Three of these units proved inoperable after deployment, shown as unshaded,
  leaving 7 non-uniformly spaced antennas for the beamforming trigger.
  Not shown here are the deployed Hpol antennas at -183~m and -185~m, 
  which are not part of the beamforming trigger.}
\label{fig:trigger_array}
\end{figure*}

The IceCube direct plane-wave impulse 
was used to determine the systematic channel-to-channel timing offsets, 
which primarily originate from small length differences of individual 
fibers in the 200~m cable.
To a lesser extent, these timing offsets are 
also caused by small non-uniformities in the physical
spacing of the NuPhase antennas, to which we assign a 
$\sim$2~cm error ($\leq$100~ps) as measured during deployment.
Several thousand IceCube pulser events were recorded and the relative time-of-arrival
of the direct pulse was measured at each NuPhase Vpol channel in the array using the cross-correlation method described above.
Assuming a plane wave impulse, the per-channel timing offsets are 
given by the residuals of a linear fit to the relative arrival times versus the antenna positions.
The channel-to-channel timing mismatches are found to be 
in the 100-400~ps range, smaller than the sampling time
resolution of the ADC. 

The relative time-of-arrival of pulses from the ARA5 calibration pulser
is shown in Fig.~\ref{fig:timing}, 
in which we plot the measured times pre- and
post-correction of the channel timing offsets. 
The time corrections shown in Fig.~\ref{fig:timing}
are applied only at the software level.
In the current trigger implementation, we don not apply a real-time correction
for these timing offsets in the beamforming firmware\footnote{Implementing 
up-sampling or fractional-delay filters on the FPGA
would allow the correction of these sub-sample offsets, but would 
add latency to the trigger output.}, 
which somewhat reduces the trigger sensitivity 
as discussed in Sec.~\ref{sec:trigger}.

\section{The Beamforming Trigger}
\label{sec:beamformingfirm} 


The digitized signals are split within the FPGA: the trigger path
sends data to the beamformer and the recording path sends data through
a programmable pre-trigger delay buffer to
random-access memory blocks on the device.
The FPGA beamforming module operates on the lower 5 bits, so that
the coherent sum does not exceed an 8-bit value. 
The RF signal level is balanced between channels 
using the digital attenuator (shown in Fig.~\ref{fig:signal_chain}) such
that the RMS voltage noise level is resolved at between 2.5 and 3 bits.
If a signal exceeds the 5-bit level, 
the trigger-path sample is re-assigned the maximum or minimum value ($\pm$15 ADC counts = $\pm$109~mV).

A schematic of the array of Vpol antennas is shown in Fig.~\ref{fig:trigger_array}. 
Three of the ten deployed Vpol antennas were non-functional after deployment,
likely caused by breaks in the mechanical-electrical connections 
at the antenna feedpoint caused while lowering the string in the borehole.
The beamformer operates on the 7 working channels, which have a non-uniform spacing.

Our beamforming trigger strategy is to form the coherent sums 
using the highest possible number
of antennas in the array (smallest baseline) as these provide the greatest SNR boost.
Coherent sums made from fewer antennas (larger baselines) are included as needed until the angular range of interest is adequately covered.
In the NuPhase beamformer, we target an elevation angle range of $\pm$50$^{\circ}$ where the Vpol birdcage antennas have good response.

In order to cover a $\sim$100$^{\circ}$ span of elevation angles, 
two sets of coherent sums are formed in the NuPhase system: one using 7 antennas with 1~m
baseline spacing (V1,3,5,6,7,8, and 9) in Fig~\ref{fig:trigger_array}.
and the other using 5 antennas with 2~m baseline spacing (V1,3,5,7, and 9)\footnote{The 
original plan was to beamform the central 8 Vpol antennas: V2-9 shown in Fig.~\ref{fig:trigger_array}.
The strategy for this 8-antenna beamformer was to use the 8-antenna
1~m baseline coherent sum in combination with a pair of 4-antenna 2~m
baseline coherent sums. This provided both an additional antenna and
a more compact array (better angular coverage) compared to the as-implemented 
trigger, which is constrained by the number of working antennas.}.
A 3~m baseline coherent sum is possible using antennas V3,6,9 and,
at longer baselines ($\geq$4~m), only pairs of antennas can be coherently summed.
These coherent sums do not add significant contributions to the trigger.

The coherent sums are calculated using
\begin{equation}
  \begin{aligned}
  S_{m}(t) = \sum_{j}^{N_{ant}} V_{j}(t - n_{m,j} \: \Delta{t} ) ,\\
  \end{aligned}
\end{equation}
where $m$ is the beam number, 
$\Delta{t}$ is the ADC sampling interval ($\sim$0.67~ns),
$V_{j}$ is the 5-bit antenna signal, and
$n_{m,j}$ is an integer that defines a beam- and antenna-specific delay. 
To fill the range of elevation angles, 15 coherent sums are simultaneously 
formed for both $N_{ant}$=5~and~7. 
The beam number, $m$, takes on a similar definition as 
introduced in Eqn.~\ref{eq:digitalbeam}, which can be used 
to calculate the adjacent beam-to-beam angular spacing.

At 180~m depth, the full NuPhase array is below the Antarctic firn layer
and embedded in deep ice, which has a relatively constant
index of refraction of $\sim$1.78.
For the $N_{ant}$=7 beams, the beam-to-beam spacing is
given by Eqn.~\ref{eq:digitalbeam}, using 
$d$=1~m and $c$=$c_{light}/1.78$, to be $\sim$6.5$^{\circ}$.
The $N_{ant}$=5 beams have a beam-to-beam spacing of $\sim$3.2$^{\circ}$.
The $N_{ant}$=5 beams that overlap with the $N_{ant}$=7 beams are not formed,
as they are redundant.

A proxy for the beam power is calculated by simply squaring
each sample in the 8-bit coherent sum. 
Next, this `power' is summed every two samples ($\sim$1.3~ns), 
which reduces the sampling resolution 
at this stage of the trigger path.
The two-sample power sums are then 
further combined between adjacent $N_{ant}$=7 and $N_{ant}$=5 beams 
so that there are now 15 equally constituted beams, 
each an independent trigger channel corresponding to a specific incoming wave direction.
This allows each beam to be set with a comparable threshold level and reduces
the overall control and feedback required to monitor all independent beams.
At this point, the total duration of the rectangular power summing window 
can be extended up to 64-samples in length. 
We program the power-summing window to 16 samples ($\sim$10.7~ns) 
corresponding to the expected pulse 
dispersion shown in Fig.~\ref{fig:time_domain_response}. 

\begin{figure*}[]
\centering
\subfloat[]{\includegraphics[trim=0.1cm 0cm 0.6cm 0cm,height=7.95cm, clip=true]{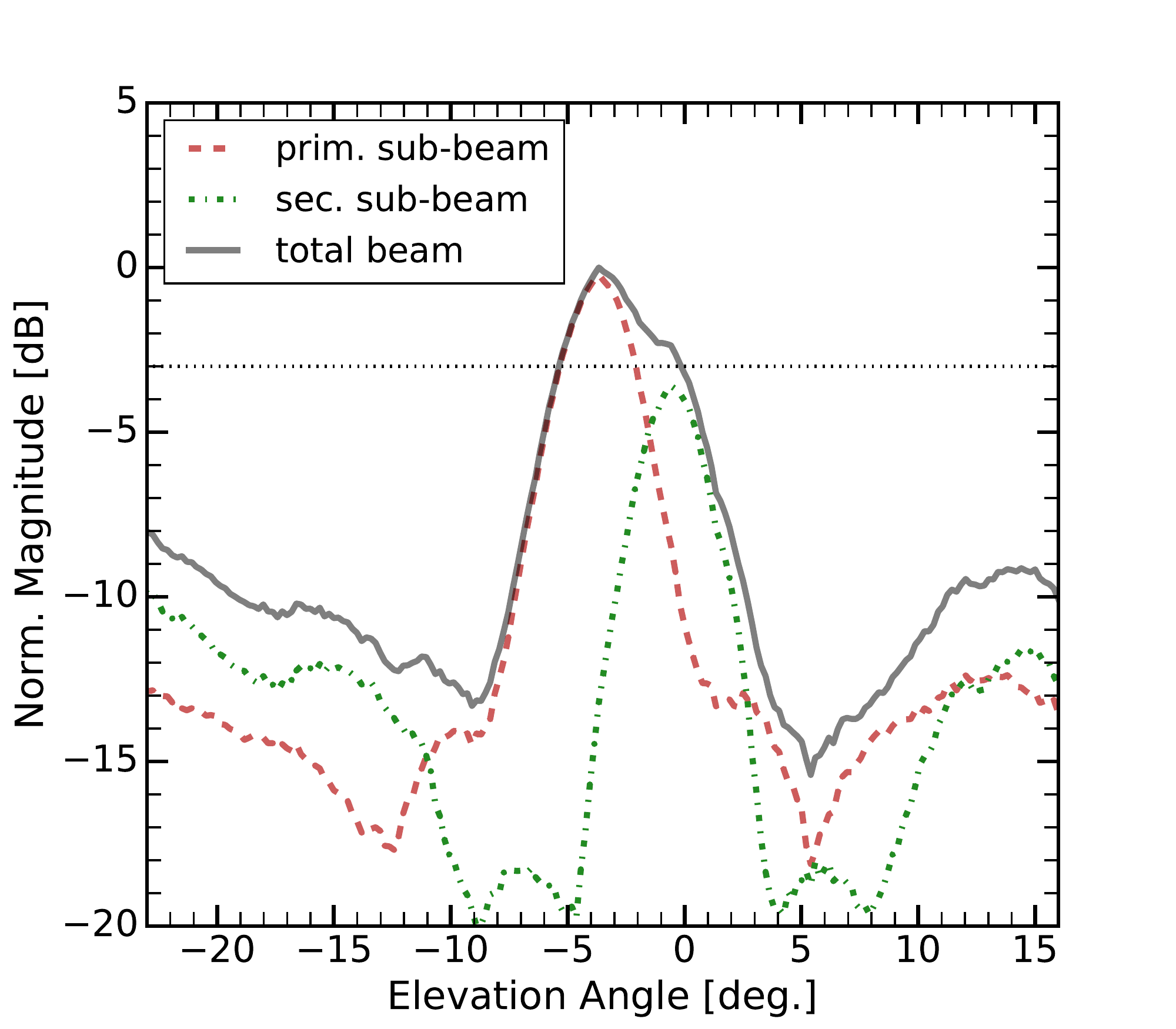}}  
\subfloat[]{\includegraphics[trim=0.1cm 0cm 0cm 0cm,height=7.95cm, clip=true]{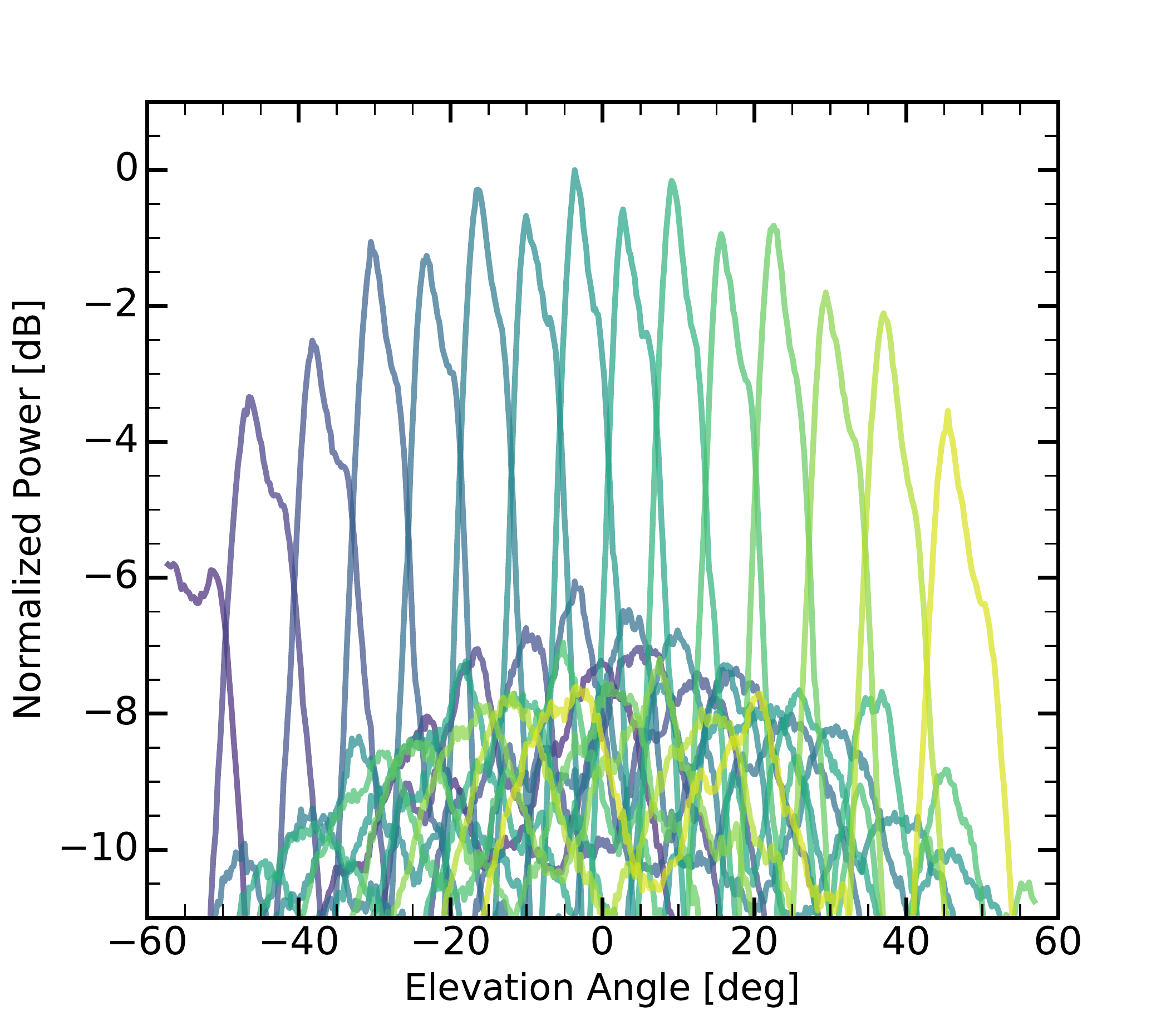}}

\caption[Simulated beam pattern, single beam]{Simulated far-field beams.
  	a) Pattern of a single NuPhase beam (beam number 7) showing constituent sub-beams, where
    the primary beam is the $N_{ant}$=7 coherent sum and the secondary beam
    is the $N_{ant}$=5 sum.
	b) All 15 beams formed on the FPGA, with each beam a separate trigger channel.
    The beams are numbered from left to right: the $m$=0 beam is centered at $\sim$-53$^{\circ}$,
    the $m$=1 beam is centered at $\sim$-45$^{\circ}$, 
    and so on, up to the $m$=14 beam  centered at $\sim$+47$^{\circ}$.
	A model for the antenna directional gain is included. 
    The beams have uniform amplitude over azimuth as given
    by the cylindrical symmetry of the birdcage antennas.
  }
\label{fig:sim_beam}
\end{figure*}

We developed a software simulation of the FPGA beamforming trigger
to optimize the coverage and understand the performance. 
A single simulated NuPhase beam is plotted on the left in Fig.~\ref{fig:sim_beam}, 
showing both the $N_{ant}$=7 (`primary') and $N_{ant}$=5 (`secondary') 
constituents using a signal-only simulation of randomly-thrown plane waves
with the system time-domain response. 
The $N_{ant}$=5 beam has a peak power about 4~dB down from the $N_{ant}$=7
beam due to having fewer antennas in its coherent sum. 
The beamwidths of each of the $N_{ant}$=7 and $N_{ant}$=5 `subbeams' are consistent 
with expectations from Eqn.~\ref{eq:beamwidth} using a 1~ns band-limited timing resolution, 
which predicts $\sim$3$^{\circ}$ and $\sim$4$^{\circ}$ FWHM beamwidth, respectively.

The resulting total beam has a FWHM beamwidth of $\sim$7$^{\circ}$.
The full 15-beam trigger coverage is shown in Fig.~\ref{fig:sim_beam}b,
which includes the Vpol antenna gain pattern. 
Each beam is an independent trigger channel that is separately thresholded. 
The NuPhase beam numbering scheme starts with the lowest pointing 
beam as $m$=0 up to the highest pointing beam, $m$=14.

\begin{figure}[]
\centering
\includegraphics[trim=0cm 1cm 0cm 1.2cm, height=11.5cm, clip=true]{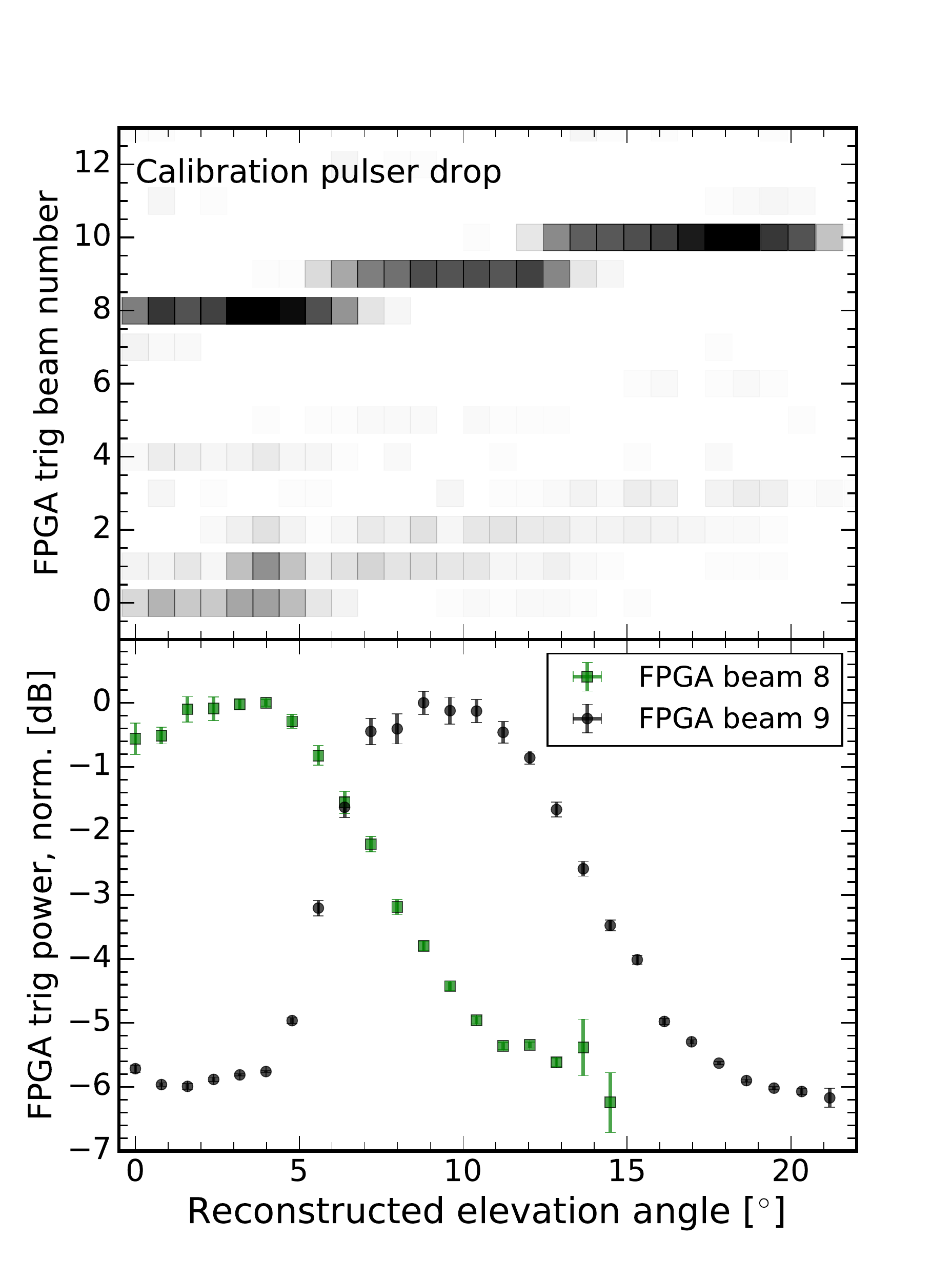}
\caption[beam mapping using pulser drop]{
	Beam mapping from a vertical scan of the ARA5 calibration pulser. 
    The top panel shows the triggered beam number as a function of the 
    reconstructed pulser location, with the marker shade indicative of 
    the number of events in each bin.
	The majority of triggers occur in beams 8, 9, and 10, 
    consistent with the vertical extent of the pulser scan.
  	A number of sidelobe triggers are also visible.
  	The bottom panel shows the triggered FPGA power for beams 8 and 9, 
    which provide a proxy for the beam pattern. Note that the beams are 
    wider than expected for far away plane waves (Fig.~\ref{fig:sim_beam}), 
    which is due to receiving the spherical wavefront shown in Fig.~\ref{fig:timing}.}
\label{fig:beam_mapping}
\end{figure}

The directional capabilities of the NuPhase trigger were tested {\it in situ}
during the deployment of the ARA5 calibration pulser string. The
Vpol transmitting antenna was enabled while lowering the calibration
string into place.
The FPGA trigger conditions, including the triggered beam number and 
calculated power, are saved with the metadata in each NuPhase event 
allowing an offline evaluation of the trigger operation.

The directional trigger response during the final $\sim$20~m 
of the Vpol pulser vertical descent is shown in Fig.~\ref{fig:beam_mapping}.
In the top panel, the FPGA triggered beam number is plotted versus
the reconstructed elevation angle. NuPhase beams 8,9, and 10 
correspond to beams centered at $\sim$3$^{\circ}$, 10$^{\circ}$, and 17$^{\circ}$, 
respectively, as shown in Fig.~\ref{fig:sim_beam}. 
A number of `sidelobe' triggers are also found in beams 0-3, 
increasing in quantity as the reconstructed angle nears 
horizontal because the pulser and receiving antennas 
become boresight-aligned (the received pulse amplitude is increased).
The elevation angle is calculated by the time difference between the 
central two Vpol antennas in the array, an approximation due to the
non-negligible spherical nature of the calibration pulser wavefront. 
The Vpol transmitting antenna was permanently installed at a depth of 
174~m, just below the top the NuPhase Vpol array as can be seen 
in Fig.~\ref{fig:trigger_array}, within the view of trigger beam number 8.

The normalized beam power in NuPhase beams 8 and 9 during the pulser drop
is shown in the bottom plot in Fig.~\ref{fig:beam_mapping}. The measured
FWHM beamwidth is 10$^{\circ}$, wider than simulated 
for the far-field response (Fig.~\ref{fig:sim_beam}), 
but understood due to beam `smearing' caused 
by the near-field calibration pulser. 
The plane wave hypothesis involved in the beamforming trigger
is non-optimal for the calibration pulser and power is spread 
among a number of adjacent beam directions.

\section{Trigger Efficiency}
\label{sec:trigger}

\begin{figure}[]
\centering
\includegraphics[trim=0cm 0.6cm 0cm 0cm, scale=.36, clip=true]{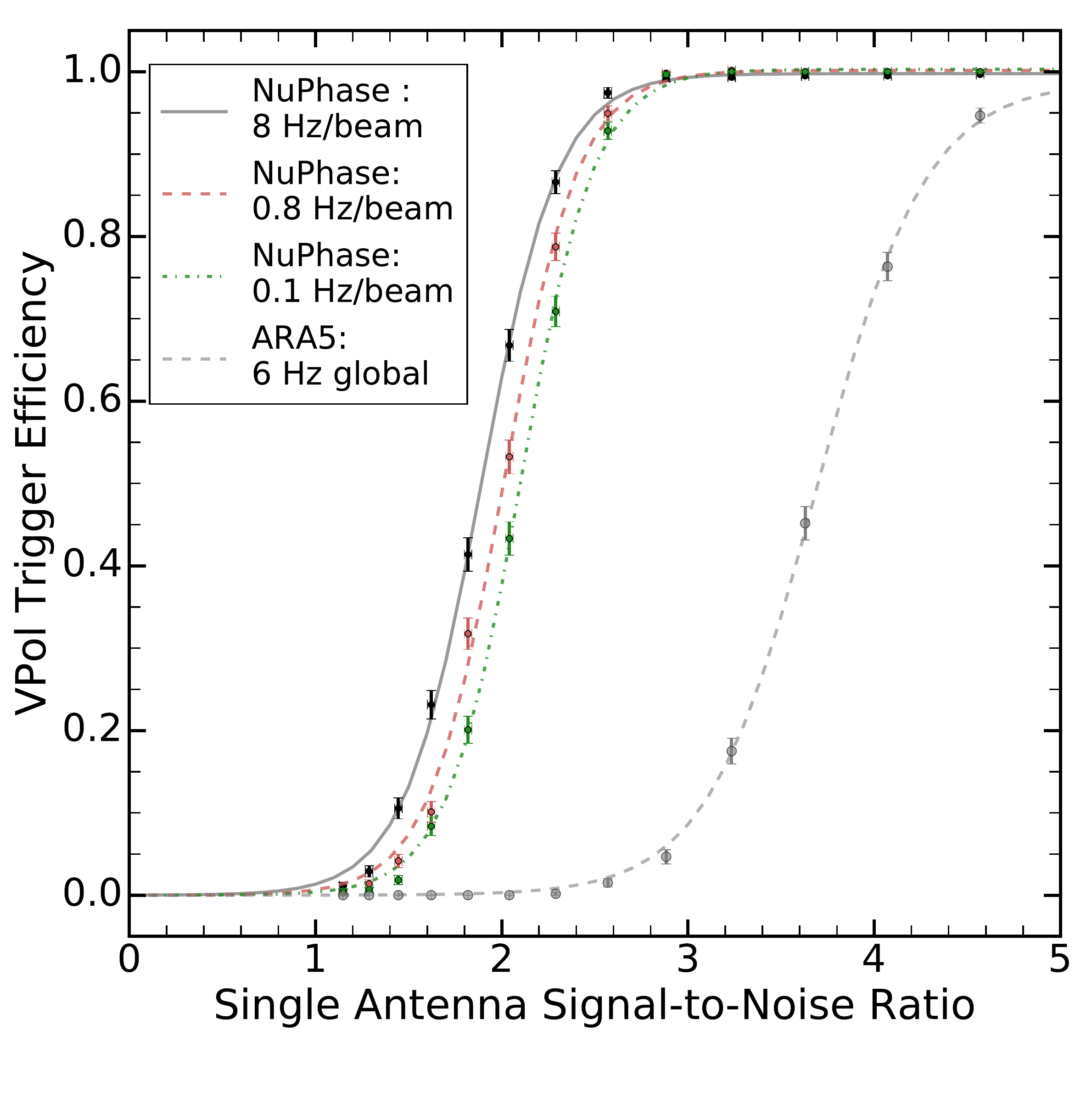}
\caption[Vpol trigger efficiency]{
	Trigger efficiency measured {\it in situ} for both NuPhase and ARA5.
  Measurements for NuPhase were taken at three different per-beam trigger rates,
  which give 50\% points at SNRs of 1.9, 2.0, and 2.1, respectively.
  ARA5 has a 50\% point at an SNR of 3.7 when operating at 6~Hz event rate.
  NuPhase uses 7~Vpol antennas in its beamforming trigger; the standard
  ARA trigger uses 8.}
\label{fig:trigger_effc}
\end{figure}

The efficiency of the NuPhase trigger was evaluated using the 
Vpol calibration pulser installed at the ARA5 station. 
The fast impulse, which is fed to the Vpol transmitting antenna,
can be attenuated in 1~dB steps, up to a maximum attenuation of 31~dB.
The calibration pulser fires at a rate of 1~Hz, timed
to the pulse-per-second (PPS) of the GPS receiver. 
To perform the measurement, pulser scans were
performed over a 10-31~dB range of attenuation, 
typically at fifteen minutes per attenuation setting.
The NuPhase trigger FPGA also receives the PPS signal where it  
is used as a $\sim$10~$\mu$s-wide gate signal that tags triggers
generated by calibration pulses, allowing a straightforward
measurement of the trigger efficiency.

The received pulse voltage SNR is defined as $V_{pp}/(2 \sigma)$
where $V_{pp}$ is the peak-to-peak signal voltage and
$\sigma$ is the voltage RMS of the thermal noise background.
The SNR is measured using the NuPhase data at each attenuation step 
in which the trigger efficiency is 100\%.
The signal $V_{pp}$ is measured by generating averaged waveforms in each
Vpol channel and taking the mean over the 7 channels. The noise RMS
is measured as an average value over the full attenuation scan.
For the high attenuation steps, where the trigger efficiency is $<$100\%, 
the real pulse SNR cannot be directly measured because the
triggered events are self-selected to be up-biased by thermal noise fluctuations. 
We therefore use a simple 1-parameter model to extrapolate 
to get the lower SNR values in the attenuation scan. 

\begin{figure}[]
\centering
\includegraphics[trim=0.9cm 0.3cm 0.5cm 1cm, height=8.75cm, clip=true]{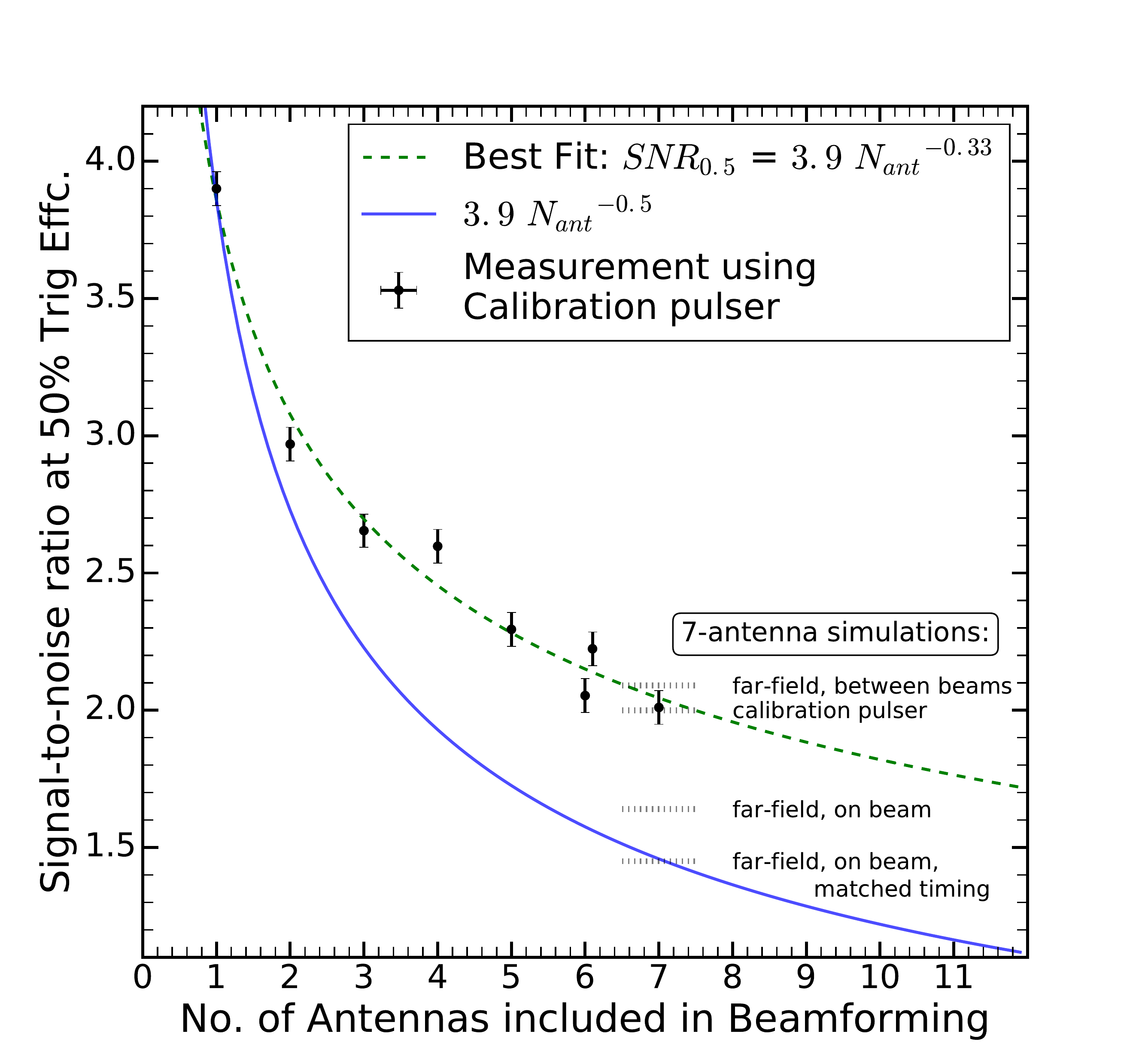}
\caption[effc vs num_antenna]{
	Trigger efficiency dependence on the number of antennas included in
  	the beamforming. The data are best fit with a scaling of $N_{ant}^{0.33}$ 
    instead of the $N_{ant}^{0.5}$ expected for coherent summing. 
    This is explained by two primary factors: 1) the spherical wave nature
  	of the calibration pulse used for the measurement, 
    and 2) systematic timing mismatches between channels.
  	Simulation results from the 7-antenna array are plotted for comparison 
    (dashed bands at the 7-antenna point), 
    which show expected efficiencies for the calibration pulser, 
    far-field on- and off-beam, and removing systematic timing mismatches 
    (shown in more detail in Fig.~\ref{fig:hardware_simulation}). 
    After accounting for the near-field nature of the calibration source, 
    beam-pattern gaps, and the timing corrections, 
    the simulation matches the $N_{ant}^{0.5}$ expectation.
    We tested two different masking configurations for the 6-antenna trigger, 
	shown by the slightly offset data points.} 

\label{fig:trigger_effc_vs_antenna}
\end{figure}

\begin{figure*}[]
  \centering
  \subfloat[]{   
  \includegraphics[trim=0.5cm 0cm 1cm 0cm,height=8.8cm, clip=true]{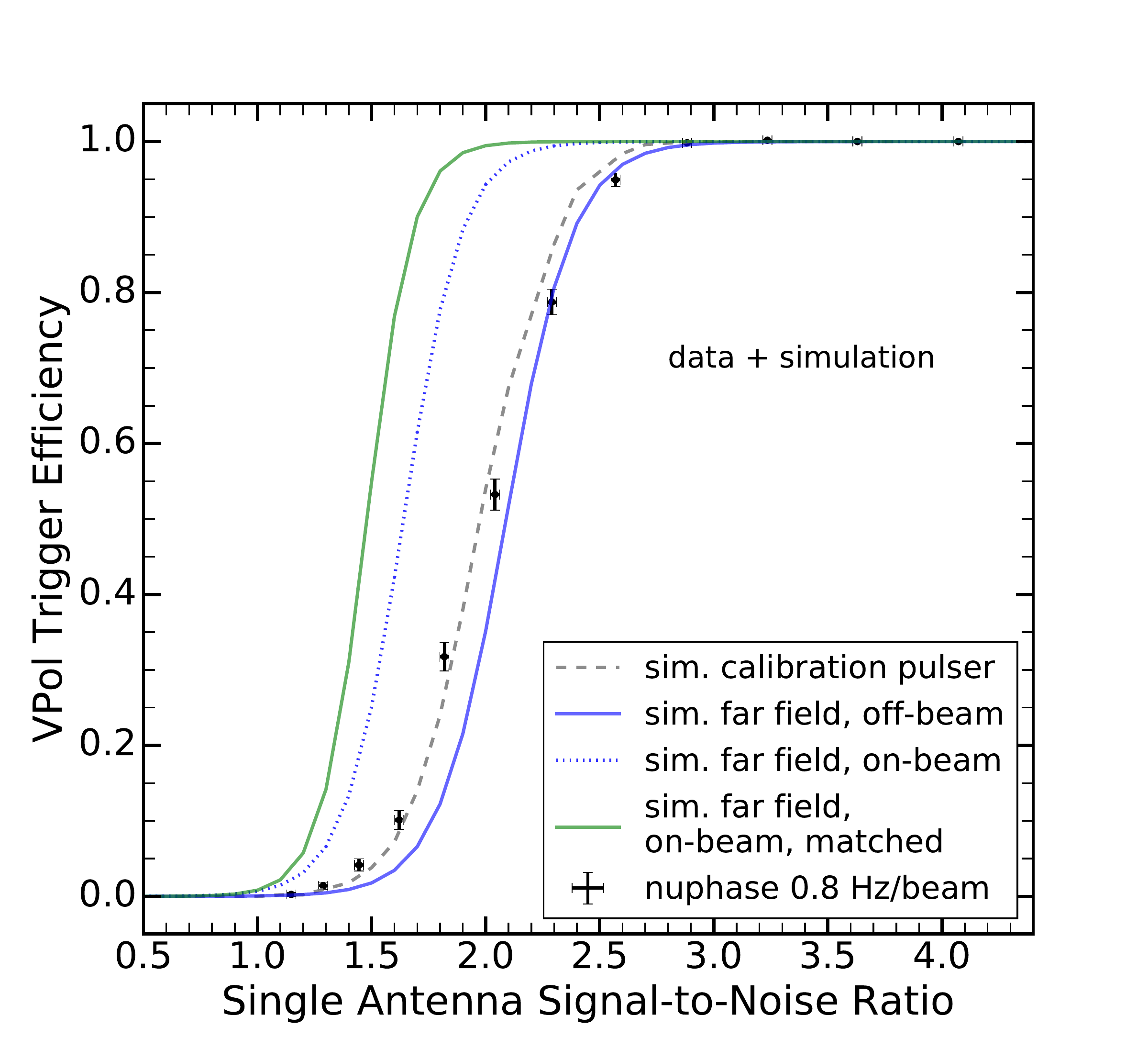}} 
  \subfloat[]{
  \includegraphics[trim=0.5cm 0cm 0cm 0cm,height=8.8cm, clip=true]{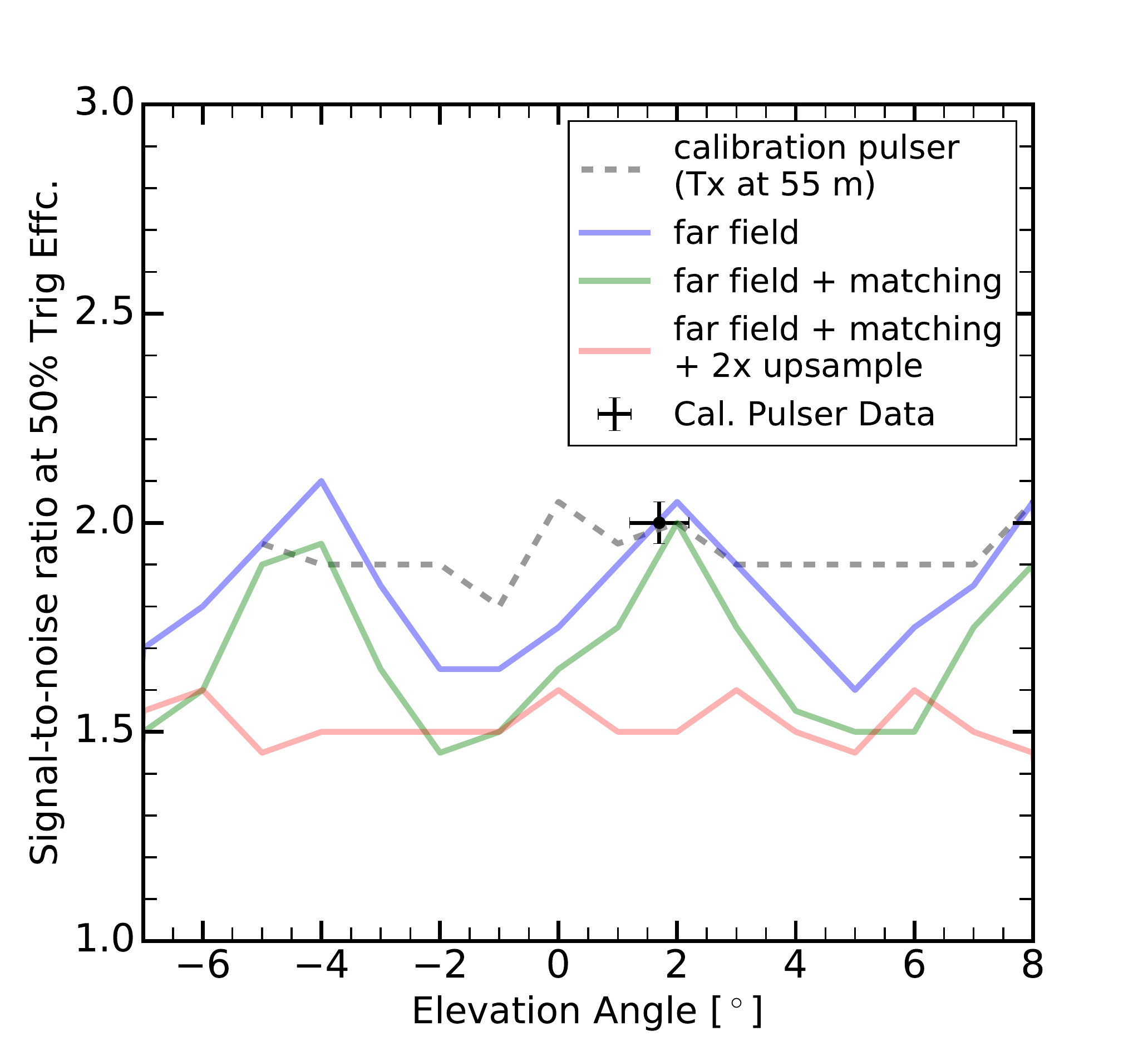}} 
  
  \caption[fpga trigger simulation]{Simulation of the hardware trigger efficiency.
    a) Four simulated efficiency curves for the NuPhase trigger:
    a local calibration pulser transmitter,
    far-field on- and off-beam, and far-field on-beam after removing timing
    mismatches between channels.
    The measurement points from the calibration pulser from
    Fig.~\ref{fig:trigger_effc} is overlaid for comparison.  
    For far-field signals, the green curve is achievable with improved timing corrections.
    b) The 50\% trigger efficiency as a function of elevation angle. 
   	The NuPhase far-field response is not uniform 
    across elevation angle due to the beamforming pattern shown
    in Fig.~\ref{fig:sim_beam}. This leads to on- and off-beam 
    efficiencies shown in (a). An `optimized' NuPhase response 
    was also simulated by removing timing mismatches
    ($\sim$8\% overall improvement) and by adding a 
    2$\times$~upsampling stage, which allows more
    delays to cover the off-beam gaps. }
\label{fig:hardware_simulation}
\end{figure*}

The measured trigger efficiency is shown in Fig.~\ref{fig:trigger_effc}.
The 50\% trigger efficiency is found at an SNR of 2.0 when running NuPhase
at a target rate of 0.75~Hz per beam for a total RF rate of $\sim$11~Hz, 
which is the nominal operation point as shown in Fig~\ref{fig:nuphase_rates}.
We also measured the trigger efficiency at lower and higher effective rates.
At 8~Hz per beam, the thresholds are set closer to the thermal noise 
background and we find a small triggering improvement with 
a 50\% point at an SNR of 1.9. 
It is not possible to run 8~Hz trigger rate simultaneously in each beam 
with the NuPhase system, so in this measurement the rate was 
kept to 0.25~Hz in the fourteen 
other beams. The trigger rate budget was essentially 
`focused' in the beam pointing towards
the calibration pulser. At a lower rate of 0.1~Hz per beam (1.5~Hz total rate), 
the 50\% point shifts to a slightly higher SNR of 2.1.

The ARA5 trigger efficiency was also measured in the pulser attenuation scans.
The 50\% trigger efficiency is found to be at an SNR of 3.7, 
similar to earlier studies shown in~\cite{araInstrument}. 
The NuPhase detector provides a factor of 1.8 lower trigger threshold in voltage
at approximately the same total trigger rate.

As discussed in Sec.~\ref{subsec:newtrigger} the sensitivity of
a coherent-summing trigger in the presence of uncorrelated noise
should improve as $N^{0.5}_{antenna}$. To test this scaling,
we ran another set of pulser attenuation scans in which we restricted
the number of channels in the NuPhase beamforming trigger. 
The measurement is shown in Fig.~\ref{fig:trigger_effc_vs_antenna}, 
in which we find the data is best fit by a $N^{0.33}_{antenna}$ scaling, 
smaller than expectations.

To understand this measurement, we added more details to the 
simulation of the FPGA trigger, including systematic timing mismatches between
channels shown in Fig.~\ref{fig:timing} and the near-field calibration pulser.
Simulated thermal noise was generated in the frequency domain by pulling random amplitudes
from a Rayleigh distribution and random phases from a uniform distribution~ for each
frequency bin~\cite{goodman}.
With the inclusion of band-matching thermal noise, we are able to recreate
the trigger efficiency that was measured using the calibration pulser,
as shown in Fig.~\ref{fig:hardware_simulation}a.

With the comparison of simulation and data, 
we find three factors contribute to the $N^{0.33}_{antenna}$ scaling:
\begin{enumerate}
\item Receiving non-plane waves from near-field calibration pulser, 
rather than a true far-field plane-wave source
\item Channel-to-channel timing mismatches
\item Beam pattern effects: the sampling rate limits the number of 
formed beams using all antennas, causing off-beam gaps
\end{enumerate}

The NuPhase far-field beam pattern, shown in Fig.~\ref{fig:sim_beam}b, is not
uniform over elevation angle and introduces an angular dependence to the
trigger efficiency as shown in Fig.~\ref{fig:hardware_simulation}b.
As currently implemented, the 50\% trigger efficiency point for far-field
plane waves varies between a highest SNR of 2.1 when then incoming plane-wave
is between beams, and a lowest SNR of 1.6 when the plane-wave is lined
up with a beam center. As shown in Fig.~\ref{fig:hardware_simulation}b,
removing the channel-to-channel timing mismatches would improve the
trigger sensitivity by 10-15\% at all elevation angles. 
The 50\% trigger efficiency points from the curves shown in Fig.~\ref{fig:hardware_simulation}a are plotted as dashed lines
at the 7-antenna point in Fig.~\ref{fig:trigger_effc_vs_antenna}.

As presented in Fig.~\ref{fig:beam_mapping}, 
we measured wider beamwidths from the calibration pulser 
vertical scan than was simulated for far-field plane waves.
When receiving spherical waves, the beams are also of smaller 
peak power and will have a corresponding drop in sensitivity.
For a nearby radio pulser, we find little angular dependence when moving its 
vertical location as shown in Fig.~\ref{fig:hardware_simulation}b. 
The trigger efficiency is shown to be roughly consistent with the `off-beam'
SNR for all angles, which is consistent with measurements.

Both the beam pattern effects and the timing mismatches could
be corrected in real-time on the FPGA, with relaxed trigger
latency requirements and sufficient FPGA resources. 
Currently, the sampling-time resolution
of the ADC ($\sim$0.67~ns) limits the ability to form more gap-filling
beams or correcting the sub-sample timing offsets.
In future implementations, this correction could be done through up-sampling 
(e.g. fractional-delay filtering or interpolation).
Fig.~\ref{fig:hardware_simulation} shows this implementation:
by correcting the time offsets and forming another set of 
FPGA beams in-between the current beams (for a total of 30 beams)
the elevation dependence is removed and the trigger efficiency
reaches a 50\% point at an SNR of 1.5 for all incoming angles. 

When these corrections are included, the 
50\% trigger efficiency point at an SNR of 1.5 
is consistent with the expected $N^{0.5}_{antenna}$
scaling for an ideal 7-antenna coherent-summing trigger, as shown 
in Fig.~\ref{fig:trigger_effc_vs_antenna}.


\section{Neutrino Simulation Studies}
\label{sec:arasim}

The NuPhase trigger performance was evaluated with ARASim, 
a Monte Carlo neutrino simulation package
developed for the ARA experiment~\cite{araTestBed,arasim}. 
The 7-Vpol antenna string of the NuPhase trigger, as shown in 
Fig.~\ref{fig:trigger_array}, was added to ARASim and time-domain waveforms
are recorded at each antenna for each simulated neutrino event.
For simplicity, the NuPhase trigger was implemented as an accept-reject algorithm 
modeled on the on- and off-beam trigger efficiency curves as shown in Fig.~\ref{fig:hardware_simulation}a 
(curves with 50\% trigger efficiencies at SNRs of 1.6 and 2.1, respectively). 
For each simulated neutrino event, the SNR is taken as the average value over 
the 7 antennas.

The effective volume of the detector, $V\Omega$, at trigger level is defined as
\begin{equation}
  V\Omega = \frac{ 4 \pi \: V_{tot}}{N_{thrown}} \: \sum_{j}^{N_{trig}}w_{j} ,
\end{equation}
where $V_{tot}$ is the physical volume in which neutrinos are thrown,
$w_{j}$ is the neutrino survival probability, and  $N_{thrown}$ and $N_{trig}$
are the number of simulated neutrinos thrown and triggered at the detector, respectively.
The effective volume was simulated from 10$^{1}$-10$^{5.5}$~PeV at 0.5 decade intervals, 
with one million neutrinos thrown at most energy steps. 
Two million events were simulated at the 
lowest two energy points to get sufficient statistics.

\begin{figure}[]
\centering
\includegraphics[trim=0.4cm 0.5cm 0.1cm 0.1cm, height=11.0cm, clip=true]{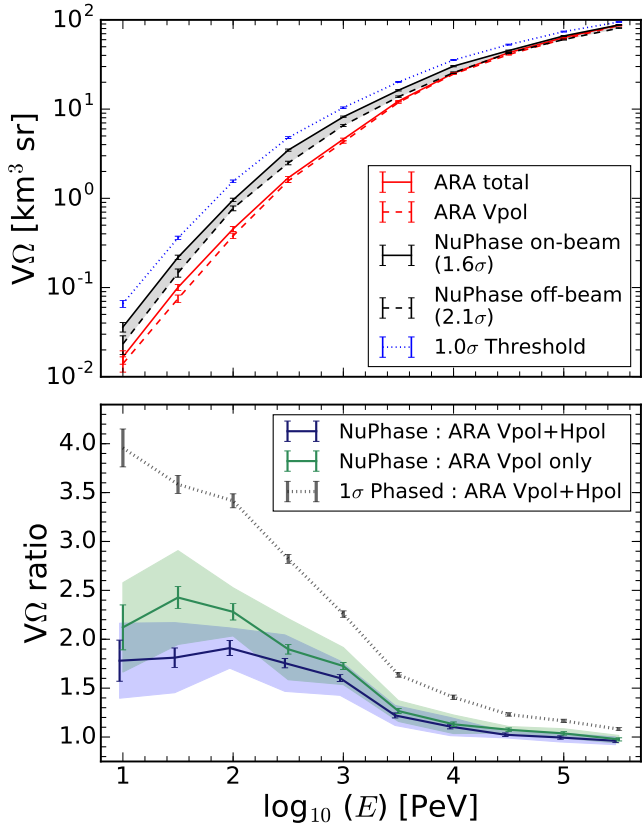}
\caption[v_{eff}]{
Trigger-level effective volume of a single-station 
ARA detector with the standard ARA trigger 
and the NuPhase trigger as simulated with ARASim.
The top panel shows the simulated effective volume 
in km$^{3}$ sr for the standard and beamforming trigger.
The solid red line is for the standard ARA dual-polarization combinatoric trigger, the
dashed red line is for a Vpol-only combinatoric trigger, the solid (dashed) black line is for the achieved 
NuPhase far-field performance maximally on- (off-) beam, and the dashed blue line
is an achievable near-term threshold with a 16-channel Vpol-only phased trigger.
The bottom panel shows the effective volume ratio of 
the beamforming trigger compared to the standard ARA trigger,
simulated as both Vpol-only and as combined Hpol+Vpol. 
The curves take into account the NuPhase beam pattern effects by averaging the off- and 
on-beam effective volumes, which are given by the solid-color bands. 
The average achieved NuPhase sensitivity compared to the standard dual-polarization
ARA combinatoric trigger and a Vpol-only ARA combinatoric trigger is shown with blue and green, respectively.
The high (low) side of the colored bands assumes the on- (off-) beam effective volume.
We also show the improvement compared to the standard dual-polarization ARA trigger that is 
achievable with a 16-channel phased trigger with a 1$\sigma$-threshold.
}
\label{fig:veff}
\end{figure}

The simulated effective volume of the NuPhase trigger is compared to the 
standard ARA combinatoric trigger in Fig.~\ref{fig:veff} for a single ARA station.
The lower panel shows the effective volume ratio between the NuPhase Vpol trigger
and two versions of the ARA trigger: 
the full dual-polarization trigger and an isolated Vpol-only trigger.
The ratios are plotted using an average of effective volumes generated
using the on- and off-beam simulated efficiency curves.
At lower energies ($\leq$300~PeV), we find the 
Vpol-only beamforming trigger increases the
ARA effective detector volume by a factor of 1.8 when 
compared to the standard ARA dual-polarization trigger.
Similarly, we find an average improvement of over a factor of 2 when comparing the 
beamforming trigger to the Vpol-only restricted ARA trigger. 
As the beamformed antennas are all Vpol, the NuPhase trigger will be blind
to events that are primarily Hpol at the ARA detector. 

The solid-color bands in the ratio plot in Fig.~\ref{fig:veff} 
show the effective volume difference between the on-beam  and off-beam 
trigger efficiency curves shown in Fig.~\ref{fig:hardware_simulation}.
These bands get wider as the neutrino energy decreases, indicating a steep
detector volume vs. trigger threshold effect at lower energies
(i.e. lower energy neutrinos will be found near threshold). 
This motivates future work to remove the off-beam gaps that produce
non-optimal trigger efficiency, which
can be done via an upsampling stage as discussed in Sec.~\ref{sec:trigger} and shown in Fig.~\ref{fig:hardware_simulation}b. 

Using the measurements shown in Fig.~\ref{fig:trigger_effc_vs_antenna}, we can 
predict the performance of a larger trigger array. Though an infinitely large
array is not possible due to the finite extent of the Askaryan signal, 
a 16-Vpol array with 1~m spacing is possible in the near term, and is only 6~m longer than 
the extent of the as-deployed NuPhase Vpol array. With the inclusion of upsampling 
to match channel-to-channel timing and to fill the elevation with 
sufficient beamforming, we can use the $N_{antenna}^{0.5}$ scaling factor. 
With this scaling, we will expect a trigger threshold at a SNR of 
$\sim$1.0 with a 16-antenna trigger array. A 1.0$\sigma$ step-function
trigger response was implemented in ARASim and the result is included
in Fig.~\ref{fig:veff}, which would result in a 3-fold increase in the 
effective volume of a single ARA station at energies $\leq$300~PeV.

\subsection{Triggered Neutrino Rates}
\begin{figure*}[]
  \centering
  \includegraphics[trim=2.5cm 0.5cm 1cm 0cm,height=10cm, clip=true]{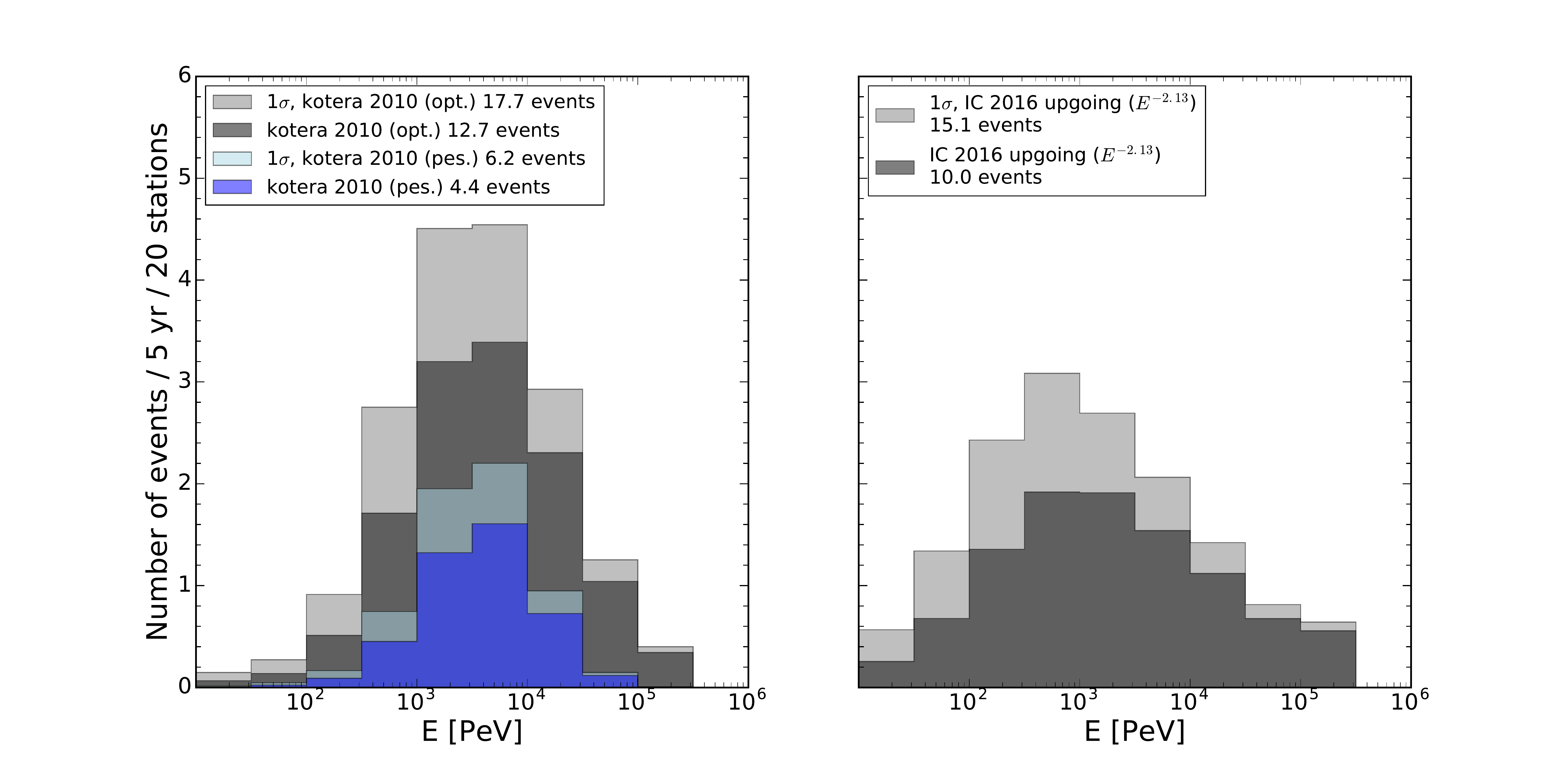} 

  \caption[neutrino rates]{
  Triggered neutrinos vs. Energy with 20 stations equipped with a 
  NuPhase trigger system in 5 years of observation.
  The triggered neutrino rate is based on the effective detector volumes using both the as-implemented 
  trigger and a improved 1$\sigma$ threshold trigger threshold.
  The left panel shows the triggered neutrino rate based on optimistic and 
  pessimistic cosmogenic fluxes~\cite{kotera}. The right panel shows the triggered 
  rate of an astrophysical flux of neutrinos based on the best-fit $E^{-2.13}$ power law from the IceCube
  up-going muon neutrino analysis~\cite{icecubeupgoing}.}
\label{fig:neutrino_rates}
\end{figure*}

Fig.~\ref{fig:neutrino_rates} shows the triggered neutrino rate for both cosmogenic and 
astrophysical flux models. These rates are calculated using the effective trigger volume 
for both the as-measured 7-channel NuPhase trigger and an improved 1$\sigma$ threshold.
The number of triggered neutrinos are shown with a 20~station detector over 5 years of observation.
For a pessimistic cosmogenic neutrino flux model, which includes no source evolution and
assumes a pure iron composition of ultra-high energy cosmic rays~\cite{kotera}, such a detector would 
capture 4.4-6.2 cosmogenic neutrinos. 
For the best-fit IceCube astrophysical flux ($E^{-2.3}$ power law) from an analysis of up-going
muon neutrinos~\cite{icecubeupgoing}, such a detector would observe 10-15.1 neutrinos, 
including 2.5-4.5 $\leq$100~PeV neutrinos.

\section{Conclusions}
\label{sec:conclude}

We describe the design and performance of a time-domain beamforming trigger
for the radio detection of high energy neutrinos. A dedicated compact array of Vpol antennas
was installed at an ARA station at South Pole in the 2017/18 season. 
Signals from these antennas are beamformed using real-time 7-bit 
digitization and FPGA processing. 
Using the ARA station near-field calibration pulser, we measure a 50\% trigger efficiency
on impulses with an SNR of~2.0. A hardware-level simulation, 
validated using calibration pulser data, predicts a 50\% trigger efficiency
on far-field (plane-wave) impulses at an SNR of 1.8$\pm$0.2. 
This SNR range is given by the realized beam pattern of trigger, which is constrained 
by the sampling time resolution of the digitized samples, 
the highest in-band frequency content of the signal, 
and the spatial extent of the antenna array.

The NuPhase triggering performance was included in the ARA neutrino simulation code, 
which shows a significant boost in the effective volume of the detector across all energies, especially large for
energies $\le10^3$~PeV. Compared to a Vpol-only ARA trigger, the already-achieved NuPhase trigger
increases the effective volume by a factor of 2 or more low energies ($\le$100~PeV).
When compared to the standard dual-polarization ARA trigger, the improvement factor drops
to an average of 1.75. 
With the addition of upsampling, which would remove the off-beam trigger efficiency 
gaps, this factor improves to 2 over the same energy range.
With the demonstrated improvement at low energies, a single ARA station is more sensitive
to a potential flux of astrophysical neutrinos. 
Using the best-fit $E^{-2.13}$ power law from as measured by IceCube~\cite{icecubeupgoing}, 
an in-ice radio detector with 20~stations equipped with the as-implemented NuPhase would detect
10 astrophysical neutrinos from this flux above 10~PeV in 5~years of observation.

Triggering algorithms with threshold-lowering potential can be tested with
the current NuPhase system by remotely re-programming the FPGA firmware.
For example, it is possible that the as-implemented 
rectangular-window power integration on the coherent sums is not optimal.
Alternative methods for setting a threshold on the coherent sums, 
such as a multiple-threshold 
requirement on the coherent sum voltage or converting the coherent 
sum to its envelope signal, may 
be better options and will be investigated. 
Finally, a real-time deconvolution of the system response in the 
FPGA would provide an increase of the SNR at trigger-level, 
improving the NuPhase performance.

We considered the possibilities of further lowering the 
trigger threshold to the 1$\sigma$
level, which would boost the effective detector volume 
by more than a factor of 3 for lower-energy ($\le$100~PeV) neutrinos.
This threshold improvement is possible with a 
16-antenna Vpol string and the addition of an upsampling block on the FPGA, but
with otherwise the same overall architecture and hardware 
of the current NuPhase trigger system.
Additionally, some combination of Hpol to and Vpol antennas in a phased trigger
may also significantly increase the sensitivity in the low-energy neutrino range, 
which is where the standard  ARA combinatoric trigger sees its largest 
fraction of Hpol-only triggered events.

\section{Acknowledgments}

We thank the National Science Foundation for their generous support through Grant NSF OPP-902483 and Grant NSF OPP-1359535. We further thank the Taiwan National Science Councils Vanguard Program: NSC 92-2628-M-002-09 and the Belgian F.R.S.-FNRS Grant4.4508.01. We are grateful to the U.S. National Science Foundation-Office of Polar Programs and the U.S. National Science Foundation-Physics Division. We also thank the University of Wisconsin Alumni Research Foundation, the University of Maryland and the Ohio State University for their support. Furthermore, we are grateful to the Raytheon Polar Services Corporation and the Antarctic Support Contractor, for field support. A. Connolly thanks the National Science Foundation for their support through CAREER award 1255557, and also the Ohio Supercomputer Center. S. A. Wissel also thanks the National Science Foundation for support under CAREER award 1752922. K. Hoffman likewise thanks the National Science Foundation for their support through CAREER award 0847658. B. A. Clark thanks the National Science Foundation for support through the Graduate Research Fellowship Program Award DGE-1343012. A. Connolly, H. Landsman, and D. Besson thank the United States-Israel Binational Science Foundation for their support through Grant 2012077. A. Connolly, A. Karle, and J. Kelley thank the National Science Foundation for the support through BIGDATA Grant 1250720. D. Besson acknowledges support from National Research Nuclear University MEPhi (Moscow Engineering Physics Institute). R. Nichol thanks the Leverhulme Trust for their support. 
This  work was  supported  by  the  Kavli  Institute  for  Cosmological  Physics  at  the  University  of  Chicago,  NSF  Award  1752922 and 1607555,  and the Sloan Foundation.
Computing resources were provided by the University of Chicago Research Computing Center.
We thank the staff of the Electronics Design Group at the University of Chicago.
We also thank L. Street and N. Scheibe for their support during the 2017/18 South Pole season.


\section*{References}
\bibliographystyle{unsrt}
\bibliography{master}


%
  


\end{document}